\documentclass[reprint,12pt]{elsarticle}
\usepackage{amssymb}
\usepackage{bm}
\usepackage{ulem}
\usepackage{xcolor}
\usepackage{lineno}

\begin{document}

\begin{frontmatter}

\title{Numerical simulation of self-oscillating catalytic reaction in a plug-flow reactor}
 
\author[a]{N.V. Peskov}\corref{*} 
\ead{peskov@cs.msu.ru}
\cortext[*]{Corresponding author}
\author[b]{M.M. Slinko}
%\author{T.M. Lysak} 
\address[a]{Faculty of Computational Mathematics and Cybernetics, \\Lomonosov Moscow State University,  Moscow, Russian Federation.}
\address[b]{Semenov Institute of Chemical Physics, Moscow, Russian Federation.}

\begin{abstract}
Self-oscillations in some oxidation reactions on metal catalysts are associated with periodic oxidation/reduction of the catalyst surface. If the reaction proceeds in a flow reactor at atmospheric pressure, then the reaction dynamics is complicated by the mass transfer in the gas phase. In this paper, the reaction in a flow reactor is described by an 1D system of advection-diffusion-reaction equations with an abstract STM-type kinetic model of the surface catalytic reaction. The dependencies of the solution of the system on the reaction parameters, the intensity of diffusion in the adsorption layer, and on the gas flow rate are studied numerically. Solutions such as localized oscillations ("breathing" structures) and mobile oxidation/reduction fronts on the catalyst surface are constructed.

\end{abstract}


\end{frontmatter}
%\linenumbers

\section{Introduction}

Heterogeneous catalytic reactions are highly non-linear systems that can operate far from thermodynamic equilibrium. They are known to exhibit complex temporal and spatiotemporal behavior including self-sustained oscillations, chemical waves and deterministic chaos \cite{1,2,3}. Spatiotemporal pattern formation has been observed both under UHV conditions over single crystal surfaces \cite{4} and under atmospheric pressure conditions on pure and supported metallic surfaces \cite{5}. To image pattern formation on surfaces under UHV conditions the photoelectron emission microscopy (PEEM) and later low energy electron microscopy (LEEM) were developed \cite{6,7}. Under atmospheric conditions infrared video thermography (IRT) has been widely used to detect and measure temperature patterns in the oxidation of CO and ethylene on a Rh/SiO$_2$ catalyst \cite{8}, oxidation of NH$_3$ on a Pt ribbon \cite{9}, CO oxidation over Pd/Al$_2$O$_3$ \cite{10}, H$_2$ oxidation on a Pt/SiO$_2$ catalyst \cite{11} and a Ni foil \cite{12}.

Under normal pressure conditions external transport limitations can significantly influence the reaction rate and the dynamic behavior of the system. In most cases IRT clearly showed that certain areas of the catalyst were hotter than others. The temperature difference between the cold and hot regions was up to 54 ℃ during H$_2$ oxidation on a Pt/SiO$_2$ \cite{10}, 150 ℃ for CO oxidation over supported Pd \cite{10} and 190 ℃ during CO oxidation over Rh/SiO$_2$ catalyst \cite{8}. That is why first simulations of spatiotemporal temperature patterns used generic kinetics, like first order exothermic reaction coupled with a slow change in surface activity \cite{13}, or polynomial kinetics \cite{14,15}, that were claimed to have similar properties to a detailed model.

Later spatiotemporal patterns during catalytic oxidation of CO over Pd supported on a glass-fiber disk-shaped cloth in a continuous reactor with feed flowing perpendicular to and through the disk were reported \cite{16}. Breathing pattern were detected representing a hot spot that expanded and contracted continuously causing the oscillations in exit CO$_2$ concentration. The temperature gradient between the hot and cold domains could be relatively small reaching only few degrees at low gas feed and low CO concentration (0.5 vol. \%). Therefore, simulations were carried out assuming that the oscillatory behavior was driven by kinetic effects which were reflected by temperature oscillations \cite{17}. The model was based on the oscillatory kinetics model of the Sales-Turner- Maple (STM) type. 

The originally suggested STM model assumes that the reaction of CO oxidation proceeds according to the Langmuir-Hinshelwood mechanism and includes the following elementary steps \cite{18}: dissociative adsorption of oxygen, O$_2$; adsorption and desorption of carbon monoxide, CO; surface reaction of CO oxidation, CO + O → CO$_2$, and dissolution and segregation of adsorbed atoms of oxygen, O. It is also suggested that dissolved oxygen blocks the O$_2$ adsorption and thus creates a feedback mechanism necessary for the origin of reaction rate self-sustained oscillations. The dissolution and segregation of oxygen can be interpreted as the oxidation and reduction of the catalyst surface. In Ref. 17 the STM type model was coupled with an enthalpy and gas-phase balances in a CSTR reactor. Although the authors did not aim to carry out a comprehensive parameter adjustment of the model, the distributed model simulated experimentally detected breathing patterns. It was demonstrated that the counteracting interaction of the reaction kinetics which tried to enlarge the hot domain and the effect of the cold boundary could lead to the formation of breathing patterns \cite{1}.

Recently it was found out that under atmospheric pressure conditions spatiotemporal structures on the catalyst surface during reaction could be monitored by naked eye and recorded by a video camera \cite{19,20}. It was shown that such method was successful for methane, ethane, and CO oxidation over Ni and Co metallic catalysts, due the different colors of metallic and oxidic states \cite{21,22,23}. It was shown that in all reactions the propagation of oxidation (dark color) and reduction (light color) fronts was relevant in the oscillatory phenomena. The most active state was the metallic surface, while the oxidic state was less reactive. The synchronization of video observations with the in situ x-ray photoelectron spectroscopy also proved the connection of the spatiotemporal patterns during propane oxidation over Ni foil with the reversible bulk oxidation of Ni to NiO \cite{24}. Mathematical modeling of kinetic isothermal waves has practically not been carried out so far. There is only one work describing the occurrence of isothermal traveling waves of oxidation and reduction during the oxidation of CO on a nickel catalyst which arise when CO and oxygen gradients in the reactor were taken into account due to the influence of mass transfer processes on the reaction rate \cite{25}.

The main goal of this work is to study the effect of mass transfer processes on the oscillating reaction, following STM type mechanism in a plug flow reactor. Both gradients in the gas phase and catalyst surface will be considered. In order to get the most general results that are not related to any particular reaction, an abstract kinetic model will be used.

\section{Kinetic model}

The abstract model describes the reaction between an oxidizing agent X and a reducing agent Y on the surface of a metal catalyst M. The elementary stages of this reaction are shown in Table 1, where $C_x$ and $C_y$ denote the concentrations of X and Y in the gas phase, and, respectively, x and y denote adsorbed particles of the X and Y species. Designations $x$ and $y$, ($x \geq 0$, $y\geq 0$, $x + y \leq 1$) are the fraction of the catalyst surface covered with particles x and y, respectively; $z$ ($0 \leq z \leq 1$) is the fraction of the oxidized surface Mx; $k_i$ is the rate constant of corresponding stage. It should be noted that since the most common oxidizing agent is oxygen, two free active sites are necessary for X adsorption. Moreover as the oscillating mechanism of STM type will be used it is suggested, that the adsorption of X occurs only on the reduced catalyst surface.

\begin{table}[h]
\caption{Elementary stages of catalytic  reaction X + Y $\to$ XY}
\centering
%\begin{tabular}{| l | l | l |}
\begin{tabular}{| l l l |}
\hline
stage & description & stage rate \\
\hline
 & & \\
X adsorption & X + 2M $\to$ 2x & $R_1 = C_xk_1(1-x-y)^2(1-z)^2$ \\
Y adsorption & Y + M $\to$ y & $R_2 = C_yk_2(1-x-y)$ \\
y desorption & y $\to$ Y + M & $R_3 = k_3y$ \\
x + y reaction & x + y $\to$ XY + 2M & $R_4 = k_4xy$ \\
catalyst oxidation & M + x $\to$ Mx & $R_5 = k_5x(1-z)$ \\
catalyst reduction & Mx + y $\to$ M + XY & $R_6 = k_6yz$ \\
 & & \\
\hline
\end{tabular}
\end{table}

The corresponding system of kinetic equations is written as:
\begin{eqnarray}
\label{ke}
&&\dot{x} = C_xk_1(1-x-y)^2(1-z)^2 - k_4xy - k_5x(1-z), \nonumber \\
&&\dot{y} = C_yk_2(1-x-y) - k_3y - k_4xy - k_6yz, \\
&&\dot{z} = k_5x(1-z) - k_6yz. \nonumber
\end{eqnarray}

Following \cite{18}, the dimensionless time $\tau = C_xk_1t$ is introduced and the dimensionless system is obtained:
\begin{eqnarray}
\label{kedl}
&&\dot{x} = (1-x-y)^2(1-z)^2 - \kappa_4xy - \kappa_5x(1-z), \nonumber \\
&&\dot{y} = p(1-x-y) - \kappa_3y - \kappa_4xy - \kappa_6yz, \\
&&\dot{z} = \kappa_5x(1-z) - \kappa_6yz, \nonumber
\end{eqnarray}
where $p=(C_yk_2)/(C_xk_1)$ is the new parameter and $\kappa_i=k_i/(C_xk_1)$.

\begin{figure}[h]
	\centering
	\includegraphics{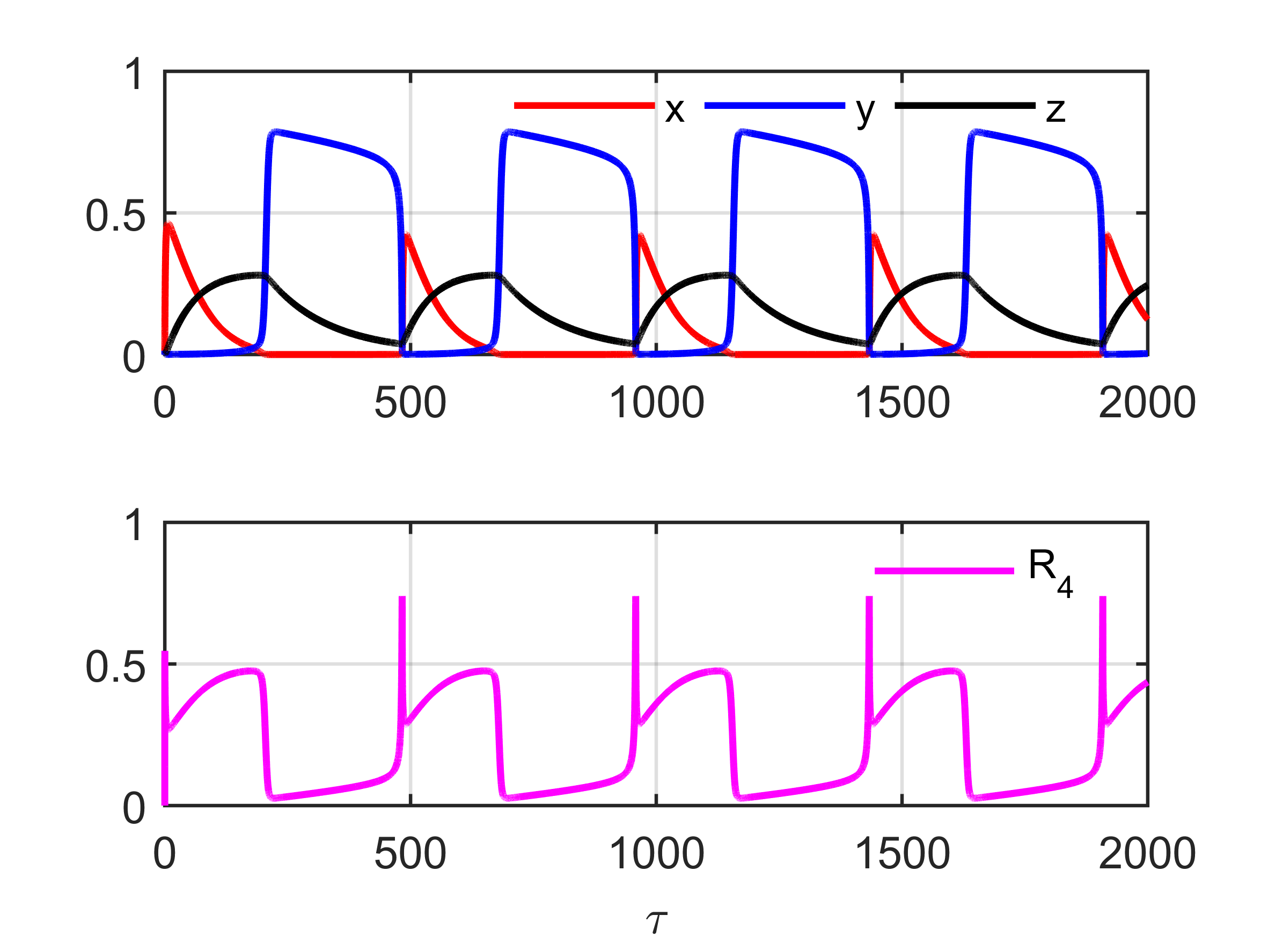}
	\caption{Top: solution of system (\ref{kedl}); bottom: corresponding reaction rate $R_4$ at $p=0.5$ and other parameters from (\ref{kap}).}
	\label{fg_km_os}
\end{figure}

Using the values of the parameters for the STM model determined in \cite{18} as the initial approximation, the following values of the dimensionless rate constants of the stages were chosen for model (\ref{kedl}):
\begin{equation}
\label{kap}
\kappa_3=10^{-1},\; \kappa_4=10^3,\; \kappa_5=\kappa_6=10^{-2}.
\end{equation}
In what follows, the parameter $p$ plays the role of an external control parameter. Figure \ref{fg_km_os} shows the solution of the system (\ref{kedl}) for $p=0.5$ starting from zero initial conditions $x(0)=y(0)=z(0)=0$.

The steady solution has the form of regular oscillations, the period of which is clearly divided into two successive parts or stages: oxidation ($\dot{z}>0$) and reduction ($\dot{z}<0$) of the catalyst. During the oxidation stage, the reaction rate is much higher than during the reduction stage. Of the rate constants (\ref{kap}), the solution is most sensitive to a change in the value of $\kappa_3$. For fixed values of the remaining parameters, self-oscillations exist in the interval $0.085<\kappa_3<0.28$. The parameters $\kappa_4,\,\kappa_5,\,\kappa_6$ can be changed by several orders of magnitude without a qualitative variation in the solution. The oscillation period is determined by constants $\kappa_5,\:\kappa_6$ and can vary by several orders of magnitude. The ratio of parameters $\kappa_5/\kappa_6$ determines the ratio of the duration of the of oxidation and reduction stages in the oscillation period.

\begin{figure}[h]
	\centering
	\includegraphics{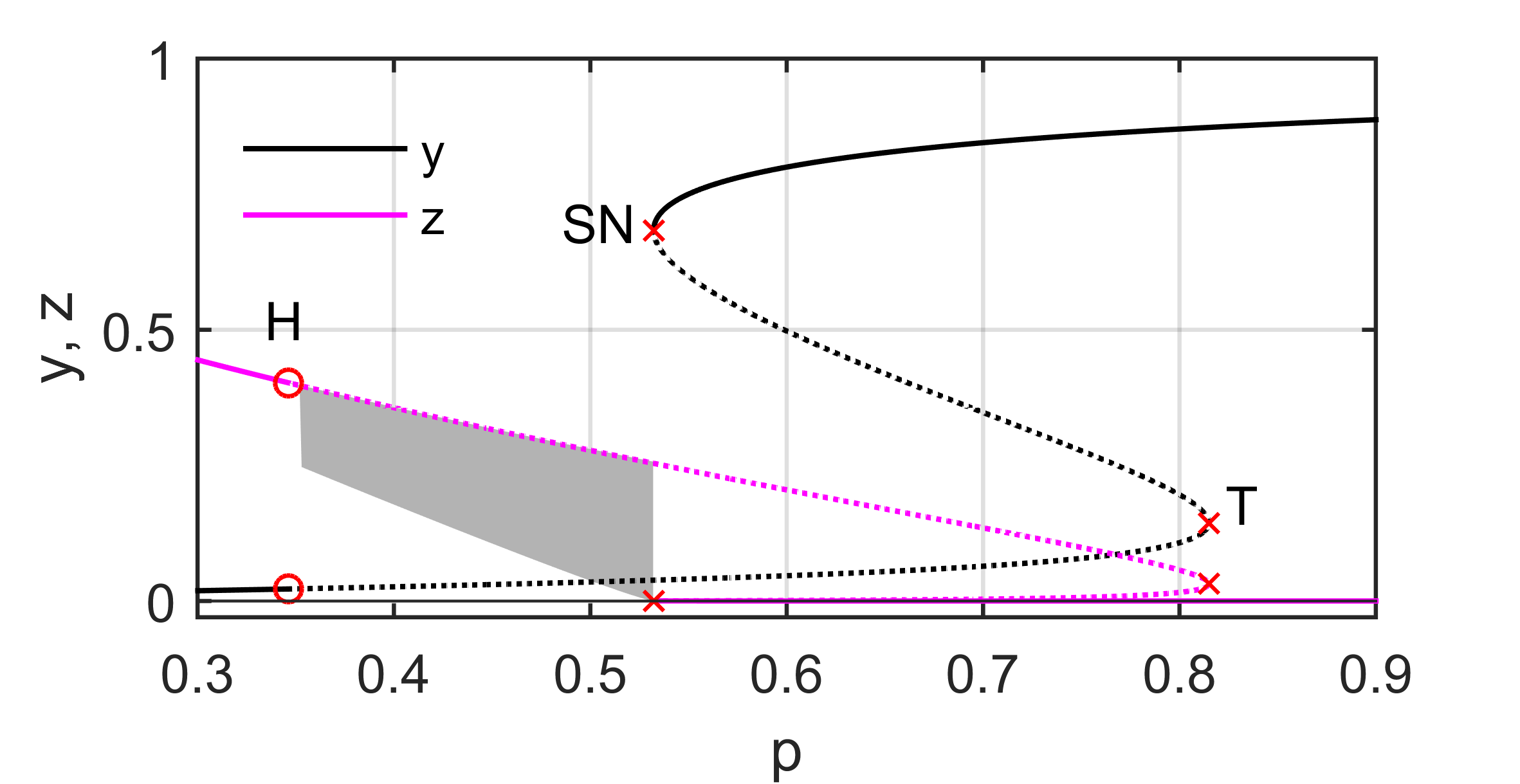}
	\caption{Bifurcation diagram of stationary solution of system (\ref{kedl}) in dependence on $p$. Solid lines -- stable solutions, dotted lines -- unstable solutions. }
	\label{fg_km_ss}
\end{figure}

Figure \ref{fg_km_ss} gives a general idea of the special solutions (attractors) of the system (\ref{kedl}) depending on the parameter $p$.

Figure \ref{fg_km_ss} shows a bifurcation diagram of a non-trivial stationary solution of system (\ref{kedl}) in dependence on $p$. In order not to obscure the picture, only the $y$ and $z$ components of the  stationary solution are shown. The stationary value of the component $x$ is close to zero in the considered range of $p$. There are three bifurcation points of the stationary solution on the interval $0< p <1$. The first point, marked `$\circ$', is the supercritical Hopf bifurcation H at $p=p_1=0.347$. At the point H, the stability of the stationary solution changes and a stable limit cycle is born. The range of oscillations in the oxidation degree $z$ in the limit cycle is shown in Figure \ref{fg_km_ss} by the gray stripe. The limit cycle vanishes in a small neighborhood of the point, marked `$\times$', of the saddle-node bifurcation SN at $p=p_2=0.532$. In this work, the type of bifurcation, in which the limit cycle disappears, and its exact position are unimportant -- one can assume that self-oscillations exist in the interval of $p$ values between the points H and SN. Passing through the point SN, three stationary states appear: one stable and two unstable. The multiplicity of stationary states exists till the third bifurcation point T at $p$ = 0.815, after which there is only one stable stationary state.

In addition to the non-trivial stationary solution discussed above, there is one trivial solution: $x = 1,\, y=0,\, z=1$, which is unstable for any value of $p$. There is no other stationary solution of system (\ref{kedl}).

\section{The model of the flow reactor}

A flow reactor is modeled as a rectangular channel with a width of $L$ m and a height of $L/2$ m. The flat square catalyst of size $L\times  L$ m$^2$ is located at the bottom of the channel. A gas mixture containing an oxidant X and a reducing agent Y flows through the channel at a linear velocity $u$ m/s.

Let $\xi$ be the coordinate along the channel in the direction of the gas flow, while the catalyst is located on the segment $0<\xi<L$. It is assumed that the distributions of reagent concentrations, $c_x(\xi,t)$ and $c_y(\xi,t)$ [mol/m$^3$], in the gas flow along the reactor are described by a system of advection-diffusion-reaction equations: 
\begin{eqnarray}
\label{adr0}
&&\frac{\partial c_x(\xi,t)}{\partial t} + u\frac{\partial c_x(\xi,t)}{\partial \xi} 
= D_x\frac{\partial^2c_x(\xi,t)}{\partial \xi^2} - q_x(\xi,t),\nonumber\\
&&\frac{\partial c_y(\xi,t)}{\partial t} + u\frac{\partial c_y(\xi,t)}{\partial \xi} 
= D_y\frac{\partial^2c_y(\xi,t)}{\partial \xi^2} - q_y(\xi,t), 
\end{eqnarray}
where $D_x$ and $D_y$ [m$^2$/s] are the diffusion coefficient of X and Y, $q_x$ and $q_y$, [mol/(m$^3\cdot$s)], describe the mass transfer between the gas phase and the catalyst.

The exchange terms, $q_x(\xi,t)$ and $q_y(\xi,t)$, can be estimated as follows. The strip of the catalyst surface with area $L\Delta\xi$, located between $\xi$ and $\xi+\Delta\xi$ across the gas flow, contains $N_cL\Delta\xi$ of active surface centers, $N_c$ [mol/m$^2$] is the surface density of active centers. Each center adsorbs $R_1\Delta t$ molecules X and $(R_2-R_3)\Delta t$ molecules Y during the time interval $\Delta t$. Therefore, the change in the concentration X in the volume of the gas phase $L\times\Delta\xi\times L/2$ above the strip is equal to $2(R_1N_cL\Delta\xi\Delta t)/(L^2\Delta\xi) $ = $2(R_1N_c/L)\Delta t$ and
\begin{equation}
\label{qq}
q_x(\xi,t)=\frac{2N_c}{L}R_1(\xi,t),\; q_y(\xi,t)=\frac{2N_c}{L}(R_2(\xi,t)-R_3(\xi,t)).
\end{equation}

To close the system (\ref{adr0}), a system of equations of the type (\ref{ke}) for distributed surface coverages $x(\xi,t)$, $y(\xi,t)$, $z(\xi,t)$, should be added to it. For the generality of the model, the mobility of adsorbed particles must be taken into account. It is assumed that the adsorbed atoms of the oxidizing agent x and reducing agent y can migrate over the catalyst surface, while the oxidized surface atoms of the catalyst Mx are immobile. This paper considers migration of the hopping type, i.e. an atom x or y can jump from a surface center below it to another free center. Thus, the intensity of migration is determined not only by the number of surface atoms, but also by the number of free centers on the surface \cite{26}.

To quantify the surface migration, let us consider three successive intervals (cells) of width $\Delta\xi$  on the $\xi$ axis with centers at the points $\xi_-=\xi-\Delta\xi$, $\xi$ and $\xi_+=\xi+\Delta\xi$. Let's assume that the frequency of jumps from one cell to the next is proportional to the product of the number of atoms in the given cell by the number of free places in the neighboring cell, divided by the distance between the cell centers. Then, the change of concentration of x atoms in the cell $\xi$ during time $\Delta t$ can be expressed as

\[\Delta(x\Delta\xi) \approx \frac{H_x}{\Delta\xi}(x_-s + x_+s - xs_- - xs_+)\Delta t,\]

where $H_x$ [m$^2$/s] is the coefficient similar to the diffusion coefficient and $s=1-x-y$ is the surface coverage of empty centers in corresponding cell. In the limit at $\Delta\xi\to 0$,   $\Delta t\to 0$, one obtains
\begin{equation}
\label{mig}
\frac{\partial x}{\partial t} = H_x\left(s\frac{\partial^2 x}{\partial \xi^2} - x\frac{\partial^2 s}{\partial \xi^2}\right) = H_x\left((1-y)\frac{\partial^2 x}{\partial \xi^2} + x\frac{\partial^2 y}{\partial \xi^2}\right).
\end{equation}
Similar expression can be formulated for y migration.

Finally, the complete model of flow reactor can be written as:

\begin{eqnarray}
\label{fr}
&&\frac{\partial c_x}{\partial t} + u\frac{\partial c_x}{\partial \xi} 
= D_x\frac{\partial^2c_x}{\partial \xi^2} - ac_xk_1(1-x-y)^2(1-z)^2,\nonumber\\
&&\frac{\partial c_y}{\partial t} + u\frac{\partial c_y}{\partial \xi} 
= D_y\frac{\partial^2c_y}{\partial \xi^2} - a[c_yk_2(1-x-y) - k_3y], \nonumber \\
&&\frac{\partial x}{\partial t} = J_x(x,y) + c_xk_1(1-x-y)^2(1-z)^2 - k_4xy - k_5x(1-z), \\
&&\frac{\partial y}{\partial t} = J_y(y,x)+c_yk_2(1-x-y) - k_3y - k_4xy - k_6yz,  \nonumber\\
&&\frac{\partial z}{\partial t} = k_5x(1-z) - k_6yz, \nonumber
\end{eqnarray}
where $a=2N_c/L$ and migratory terms of the form (\ref{mig}) are denoted by $J_x$ and $J_y$.

Considering the concentrations of reagents in the gas phase, the model (\ref{fr}) is supplemented by the following Danckwerts boundary conditions at the inflow boundary ($\xi = 0$)
\begin{equation}
\label{bc0}
u(c_x-C_x)|_{\xi=0}=D_x\frac{\partial c_x}{\partial\xi}\bigg|_{\xi=0},\;
u(c_y-C_y)|_{\xi=0}=D_y\frac{\partial c_y}{\partial\xi}\bigg|_{\xi=0};
\end{equation}
and the Neumann boundary conditions at the outflow boundary ($\xi = L$)
\begin{equation}
\label{bc1}
\frac{\partial c_x}{\partial \xi}\bigg|_{\xi=L}=0,\; 
\frac{\partial c_y}{\partial \xi}\bigg|_{\xi=L}=0.
\end{equation}
For surface coverages $x,y,z$ the Neumann boundary conditions are set both at the inlet and at the outlet of the reactor.

In dimensionless time $\tau=C_xk_1t$ and dimensionless coordinate $\zeta=\xi/L$ ($0<\zeta<1$) the corresponding system for dimensionless concentrations $v=c_x/C_x$, $w=c_y/C_y$ is as follows:

\begin{eqnarray}
\label{frdl}
&&\frac{\partial v}{\partial \tau} + \frac{1}{\tau_L}\frac{\partial v}{\partial \zeta} 
= d_x\frac{\partial^2v}{\partial \zeta^2} - a_x v(1-x-y)^2(1-z)^2,\nonumber\\
&&\frac{\partial w}{\partial \tau} + \frac{1}{\tau_L}\frac{\partial w}{\partial \zeta} 
= d_y\frac{\partial^2w}{\partial \zeta^2} - a_y \big[pw(1-x-y) - \kappa_3y\big], \nonumber \\
&&\frac{\partial x}{\partial \tau} = j_x(x,y) + v(1-x-y)^2(1-z)^2 - \kappa_4xy - \kappa_5x(1-z),\\
&&\frac{\partial y}{\partial \tau} = j_y(y,x) + pw(1-x-y) - \kappa_3y - \kappa_4xy - \kappa_6yz,  \nonumber\\
&&\frac{\partial z}{\partial \tau} = \kappa_5x(1-z) - \kappa_6yz, \nonumber
\end{eqnarray}
Here the following notations are used:
\[\tau_L = C_xk_1t_L, \; t_L = L/u;\]
\[d_x=\frac{D_x}{C_xk_1L^2}, \; d_y=\frac{D_y}{C_xk_1L^2}; \] 
\[a_x = \frac{a}{C_x}, \; a_y =\ \frac{a}{C_y};\]
\[j_x(x,y)=\frac{J_x(x,y)}{C_xk_1L^2}, \; j_y(y,x) = \frac{J_y(y,x)}{C_xk_1L^2}.\]

\section{Parameter estimates}

To carry out numerical calculations, it is necessary to set the values of the invariable parameters of the model and determine the ranges of change in the values of the variable parameters.

In this work, the values of the rate constants $\kappa_i$ (see Eq. (\ref{kap})), adsorption constants $k_{1,2}$, and the oxidant concentration $C_x$ in the inlet gas flow are fixed. The ideal gas theory is usually used to estimate the adsorption constants. From the Maxwell distribution, an estimate of the average frequency of collisions of molecules with an adsorption site (with area $1/N_c$) on the catalyst surface follows, which is used to estimate the adsorption constant
\[k_i= \frac{s_i}{N_c}\sqrt{\frac{RT}{2\pi M_i}}, \; i = 1,2,\]
where $R$ [J/(K$\cdot$mol)] is the ideal gas constant, $T$ [K] is the gas temperature,  $M_i$ [kg/mol] is the mass of 1 mol of gas and $s$ is the sticking coefficient. The calculations will use the values of the constants calculated for O$_2$ ($M_1 = 0.032$ kg/mol) and CO ($M_2 = 0.028$ kg/mol) at the gas temperature $T = 850$ K. Also, the value of $N_c=2.66\times 10^{-5}$ for Ni surface will be used. For simplicity, we assume that $s_1 = s_2 = 0.1$. Then, one has
\[k_1=7.048\times 10^5.\; k_2 = 7.534\times 10^5\;  \mathrm{m^3/(mol\times s).}\]

Calculations are carried out for a constant concentration of the oxidizing agent $C_x$, which corresponds to a partial pressure of 2 kPa in the inlet gas mixture. Using the equation of state for the ideal gas, the following value of $C_x$ concentration was calculated:
\[C_x = 2.830\times 10^{-1}\, \mathrm{mol/m^3}.\]
Therefore, the frequency of X adsorption per site can be obtained as:
\[C_xk_1 = 1.995\times 10^5\, \mathrm{s^{-1}}.\] 

The concentration of the reducing agent, $C_y$, at a given value of the parameter $p$ is determined from the equation $p=(C_yk_2)/(C_xk_1)$.

Diffusion of reagents in the gas phase reduces the stiffness of the system of equations and facilitates the numerical solving. A reasonable change in the diffusion coefficients does not have a noticeable effect on the solutions of system (\ref{fr}). Therefore, all calculations were carried out at constant and identical values of the diffusion coefficients $D_x=D_y=D= 5 \times 10^{-5}\,\mathrm{m^2/s}$. 

While the estimates of the diffusion coefficients in the gas phase of a number of species are well known, the values of the diffusion coefficients of the reactants over the catalyst surface differ significantly. Therefore, we assume $H_x = H_y = H$, and take $H = D$ as the base value. To demonstrate the dependence of the solution on $H$, its value will be decreased and increased by an order of magnitude.

The reaction dynamics in a flow reactor is greatly influenced by the residence time $t_L$. At $t \to 0$ the dynamics behavior of the reaction will approach the description by the kinetic model, while at a sufficiently large $t_L$ the reaction will stop due to a lack of reagents. In the calculations, $t_L = 0.01$ s was chosen as the main value, which at $L = 0.01$ m corresponds to the gas flow velocity $u = 1$ m/s.

The frequency of oscillations in the kinetic model is determined mainly by the oxidation-reduction rates of the catalyst. For most of the calculations associated with regular oscillations, sufficiently high redox rates are chosen ($\kappa_5 = \kappa_6=10^{-2}$), which produce a high frequency of oscillations and, consequently, reduce the computational time.
In our computer simulation, the most interesting dynamics is observed at $\kappa_5 = 10^{-5},\,10^{-6}$, which give oscillation frequencies comparable with the experimental data. Therefore, at the end of the paper, several results for such values of rates will be presented.

\section{Numerical method}

For the numerical solution of system (\ref{frdl}), the method of finite differences is used. Derivatives with respect to $\zeta$ on the right side of the system are replaced by finite differences on a uniform grid of $N$ nodes, $\zeta_k = h/2+(k-1)h$, $k = 1,2,\dots,N$, where $h=1/N$ is the grid step. The first order derivative is replaced by the backward difference, $f^\prime_\zeta\approx(f_k-f_{k-1})/h$ (1st order of accuracy), the second order derivative is approximated by the central difference, $f^{\prime\prime}_{\zeta\zeta}=(f_{k-1}-2f_k+f_{k+1})/h^2$ (2nd order of accuracy), where $f_k=f(\zeta_k)$. 

The entries $f_0$ and $f_{N+1}$ are determined using the boundary conditions (\ref{bc0}),(\ref{bc1}). From the Danckwerts conditions at $\zeta=0$ one obtains
\[u(f_0-1)=\frac{D}{Lh}(f_1-f_0).\]
From the Neumann conditions at $\zeta=0,1$ it is followed that
\[f_0=f_1,\; f_{N+1}=f_N.\]

Thus, system (\ref{frdl}) transforms into a system of $5N$ ODEs for a $5N$-dimensional vector of unknowns $q^N=\{v_k,w_k,x_k,y_k,z_k,\,k=1,2,\dots,N\}$
\begin{equation}
\label{frfd}
\dot{q}^N(\tau)=F(q^N(\tau)).
\end{equation}
The Jacobi matrix of the system (\ref{frfd}) is 13-diagonal, so it is reasonable to use the technique of sparse matrices to solve the system numerically. Moreover, it follows from our computational experience that the numerical solution of the system ceases to noticeably depend on $N$ when $N$ is not less than 1000. Consequently, all numerical solutions of system (\ref{frfd}) presented below were obtained with $N = 1000$.

To solve the system (\ref{frfd}), the Matlab `ode15s' solver was used.

\section{Results and discussion}

This section presents the results of numerical simulation of system (\ref{frfd}) for various values of the model parameters in order to demonstrate the influence of parameters on the dynamics of the model. First of all, the dependence their solution on the parameter $p$, which was a bifurcation parameter in the kinetic model, is shown. Then, the role of another external parameter, $t_L$, is explained. From the internal parameters of the model, three parameters were chosen to demonstrate the influence upon the model behavior, the surface migration coefficient $H$ and the catalyst oxidation rate constants $\kappa_5$, $\kappa_6$. In all cases, the solutions start from the "zero" initial condition $q^N(0) = {\bm 0}$.

\subsection{High frequency oscillations}

As it was noted above, the frequency of oscillations is determined mainly by the $\kappa_5, \, \kappa_6$. As a first step, we take the same values of parameters as in the kinetic model, namely $\kappa_5 = \kappa_6 = 10^{-2}$. In this case, the model (\ref{frfd}) produces oscillations with (dimensionless) period of the order of several $\tau_L$ that corresponds to the real time oscillation frequency about $10^2$ s$^{-1}$.

\subsubsection{Dependence of the solution on the parameter $p$}

\begin{figure}[h]
	\centering
	\includegraphics{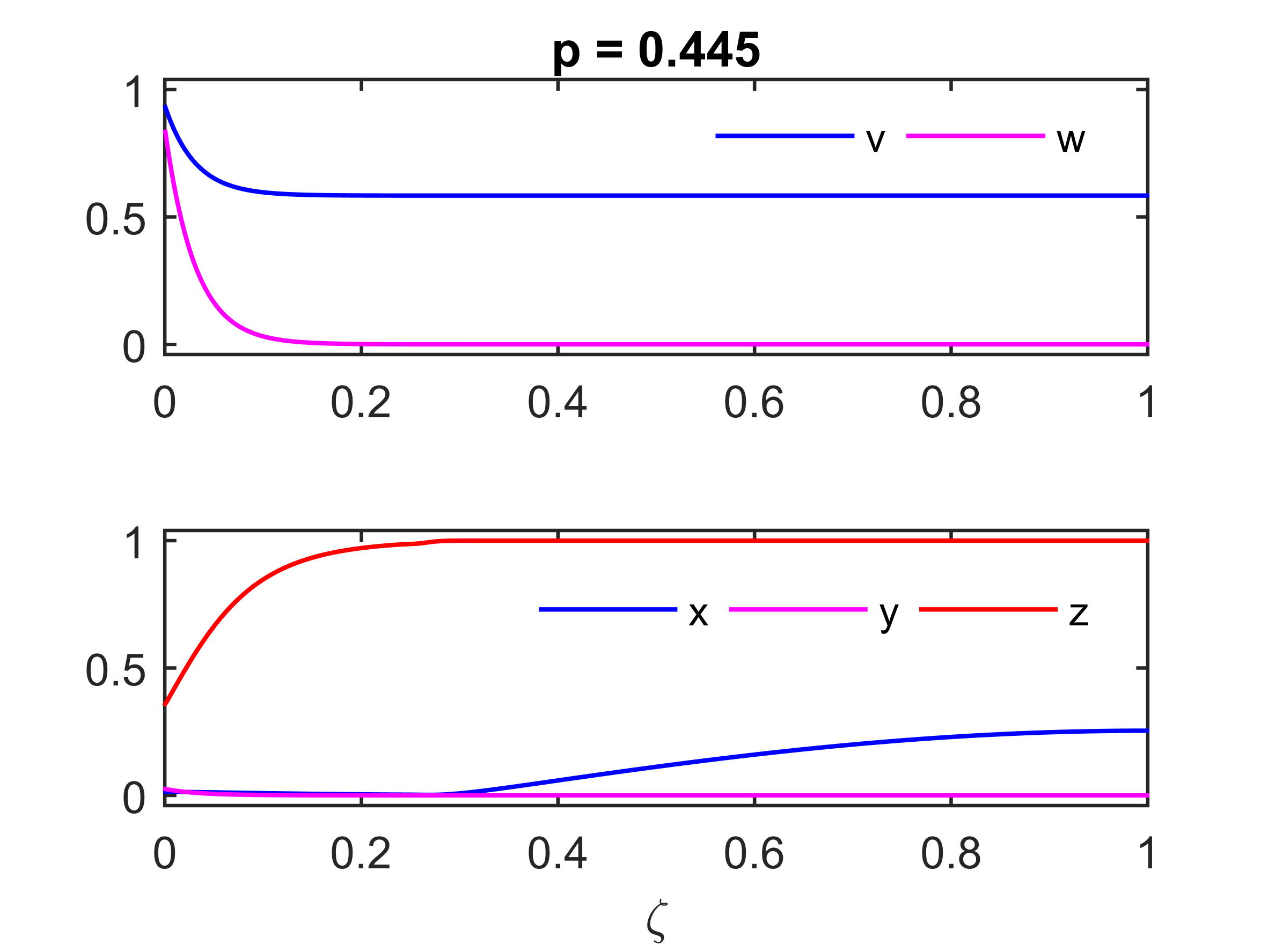}
	\caption{Stationary state of the system (\ref{frfd}) at $p=0.445$. Top: relative concentration of reagents in the gas flow. Bottom: concentrations of surface species.}
	\label{fg_ss_445}
\end{figure}

The dependence of the solution of the flow reactor model on the parameter $p$ qualitatively reproduces the dependence on this parameter of the solution of the kinetic model. Here, too, for small $p$ (more precisely, for $p < 0.446$) the solution from zero initial data tends to a stationary state. Figure \ref{fg_ss_445} shows the graphs of the stationary solution at $p = 0.445$, just before the onset of oscillations. Curve $z$ on the lower panel of Figure shows that most of the catalyst plate, with the exception of the left part (front relative to the gas flow), is completely oxidized. The reaction proceeds only in the relatively narrow frontal strip of the plate, where all the reducing agent is consumed.

For some $p$ in the interval (0,445, 0.446), the stationary solution (solution to system $F(q^N) = 0$) undergoes a Hopf bifurcation, as a result of which it becomes unstable and a stable limit cycle is born. The unstable steady state is continued numerically up to $p$ = 0.8723. With a further increase in $p$, the steady state is rapidly restructured, and the numerical solution of the equation on the uniform grid with $N = 1000$ becomes impossible.

\begin{figure}[h]
	\centering
	\includegraphics{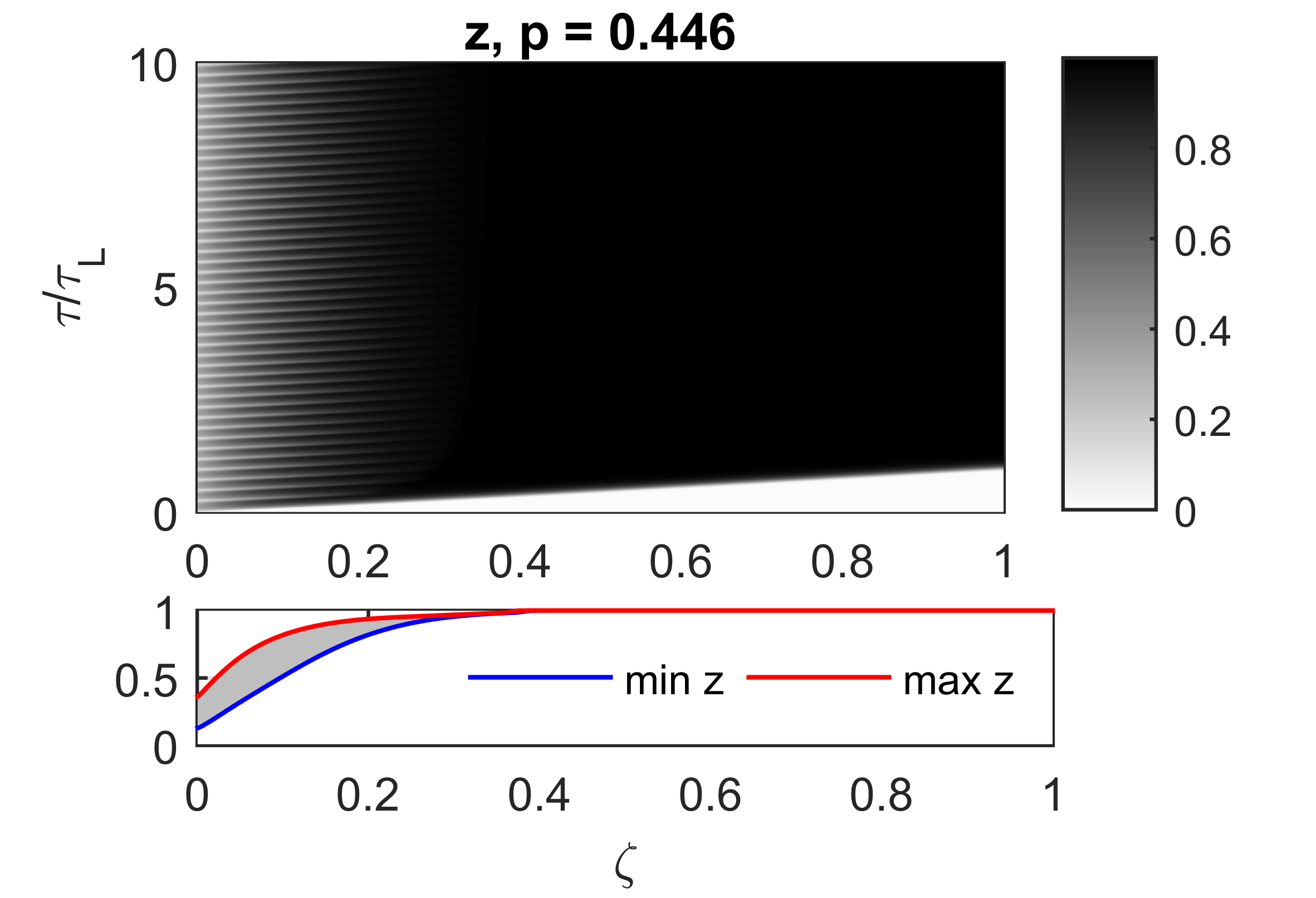}
	\caption{Top: the space-time plot of $z(\zeta.\tau)$ at $p$=0.446. Bottom: the range of $z$ oscillations in dependence on $\zeta$.}
	\label{fg_os_446}
\end{figure}

The top panel of Figure \ref{fg_os_446} shows the space-time plot of the degree of oxidation of the catalyst surface, $z(\zeta,\tau)$, at $p$ = 0.446, that is just after the onset  of the oscillations. At the front of the catalyst plate, rapid, regular oscillations in the oxidation degree are visible, which gradually decay as one moves to the right along the gas flow. The lower part of the Figure shows the range of oxidation degree oscillations depending on the dimensionless coordinate $\zeta$. Note that the oscillations have a relatively high frequency (especially in real time), which is associated with a large value of the oxidation constant $\kappa_5$ and the proximity of $p$ to the Hopf bifurcation point.

\begin{figure}[h]
	\centering
	\includegraphics{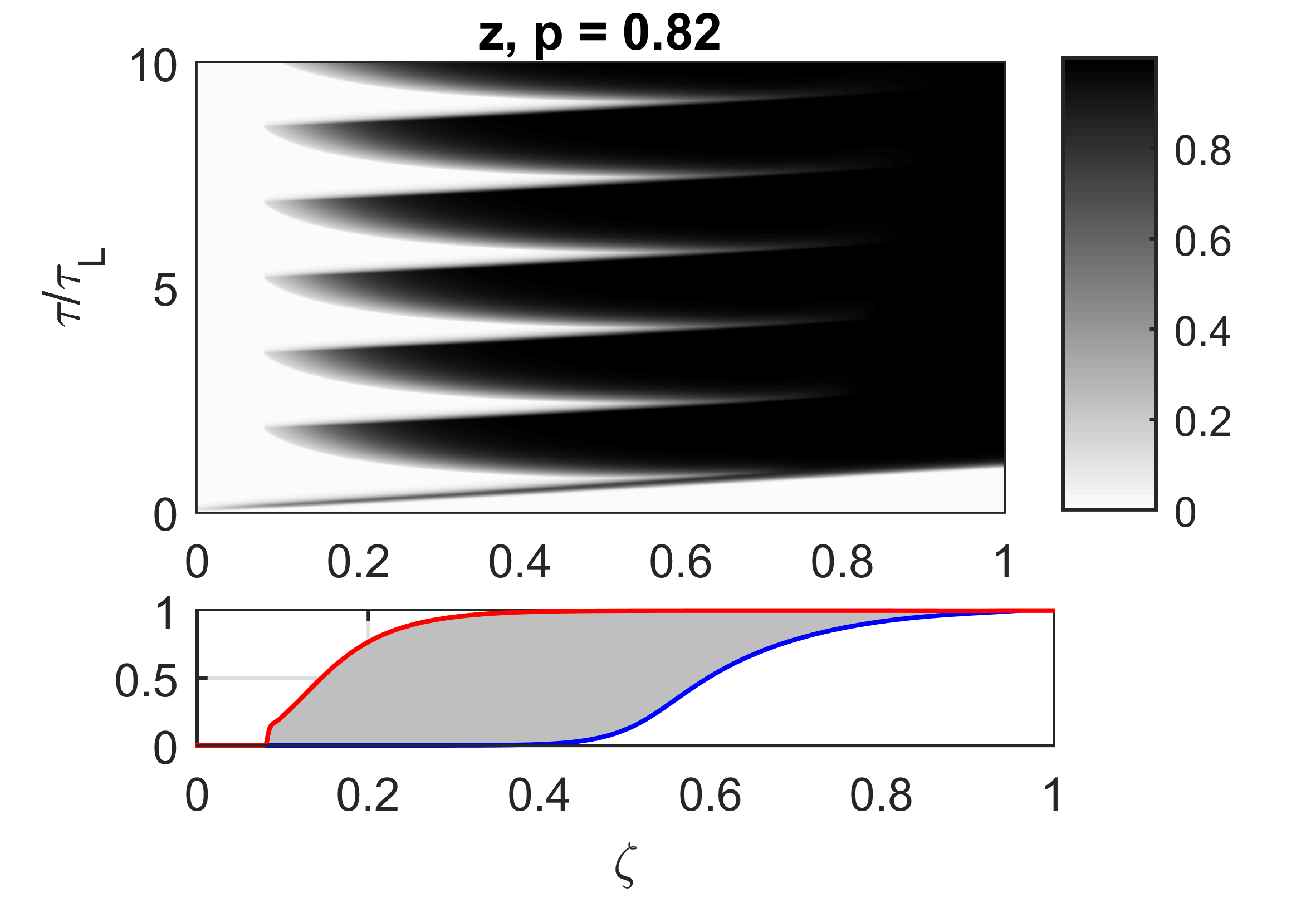}
	\caption{Same as in Figure \ref{fg_os_446} for $p$ = 0.82.}
	\label{fg_os_82}
\end{figure}

As $p$ increases, the oscillations shift deeper into the plate, their frequency decreases, and the amplitude increases. Fig. \ref{fg_os_82} shows the oscillations in the oxidation degree $z(\zeta,\tau)$ at $p$ = 0.82. In this case, the front part of the plate is constantly in a reduced state, while the back part is in an oxidized state. 
%\clearpage

\begin{figure}[h]
	\centering
	\includegraphics{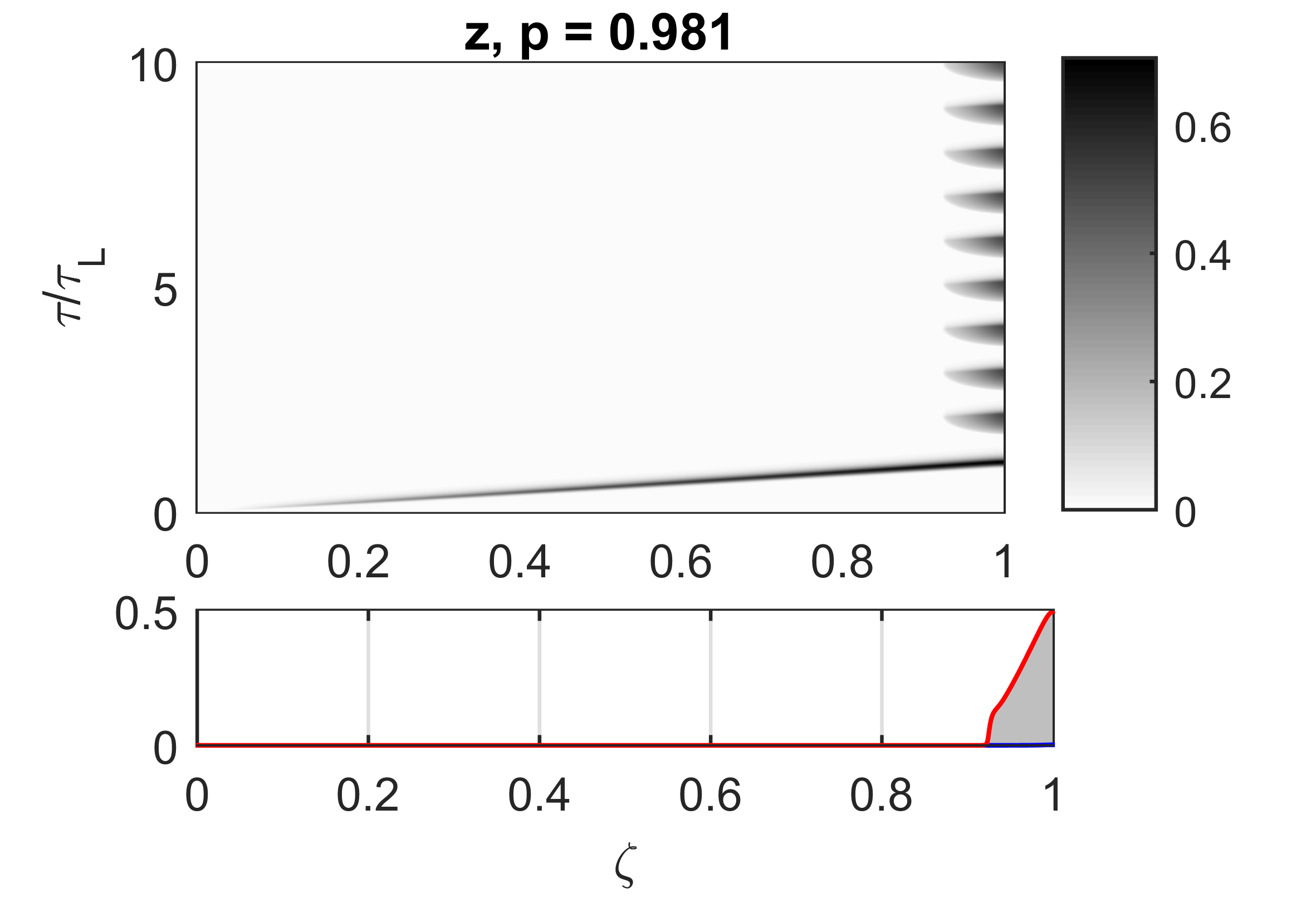}
	\caption{Same as in Figure \ref{fg_os_446} for $p$ = 0.981.}
	\label{fg_os_981}
\end{figure}

\begin{figure}[h]
	\centering
	\includegraphics{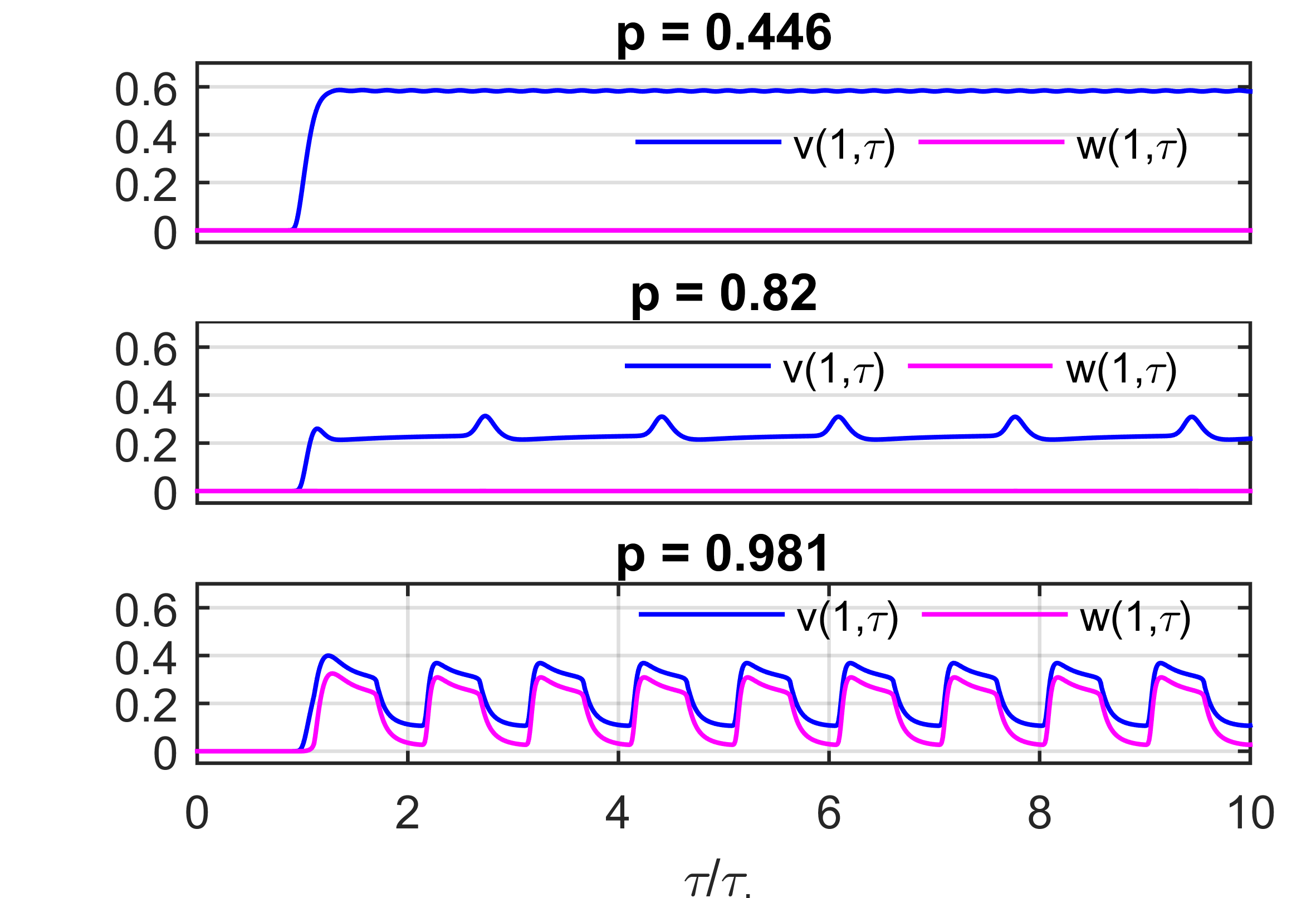}
	\caption{The concentration of reagents at the reactor outlet in dependence on time.}
	\label{fg_os_pvw}
\end{figure}

\clearpage

With a further increase in the parameter $p$, the region of oscillations moves to the right (rear) edge of the plate, as shown in Figure \ref{fg_os_981} for $p$ = 0.981. In this case, almost the entire surface of the plate is reduced. Thus, the range of values of the parameter $p$, in which oscillations exist, in the model of a flow reactor is more than 2 times wider than in the kinetic model.

Figure \ref{fg_os_pvw} shows plots of reactant concentrations $v$ and $w$ in the gas phase at the outlet of the reactor (at $\zeta=1$) as a function of time for the three values of the parameter $p$ presented in Figures \ref{fg_os_446}-\ref{fg_os_981}. When oscillations of the model variables occur far from the right edge of the plate, they are hardly noticeable in the gas flow at the outlet of the reactor. In particular, the graphs do not show oscillations in concentrations at $p$ = 0.446. And at $p$ = 0.82, despite the large amplitude of oscillations in a wide band in the central part of the plate, there are only weak fluctuations in the concentration of the oxidizing agent at the outlet flow.

\begin{figure}[h]
	\centering
	\includegraphics{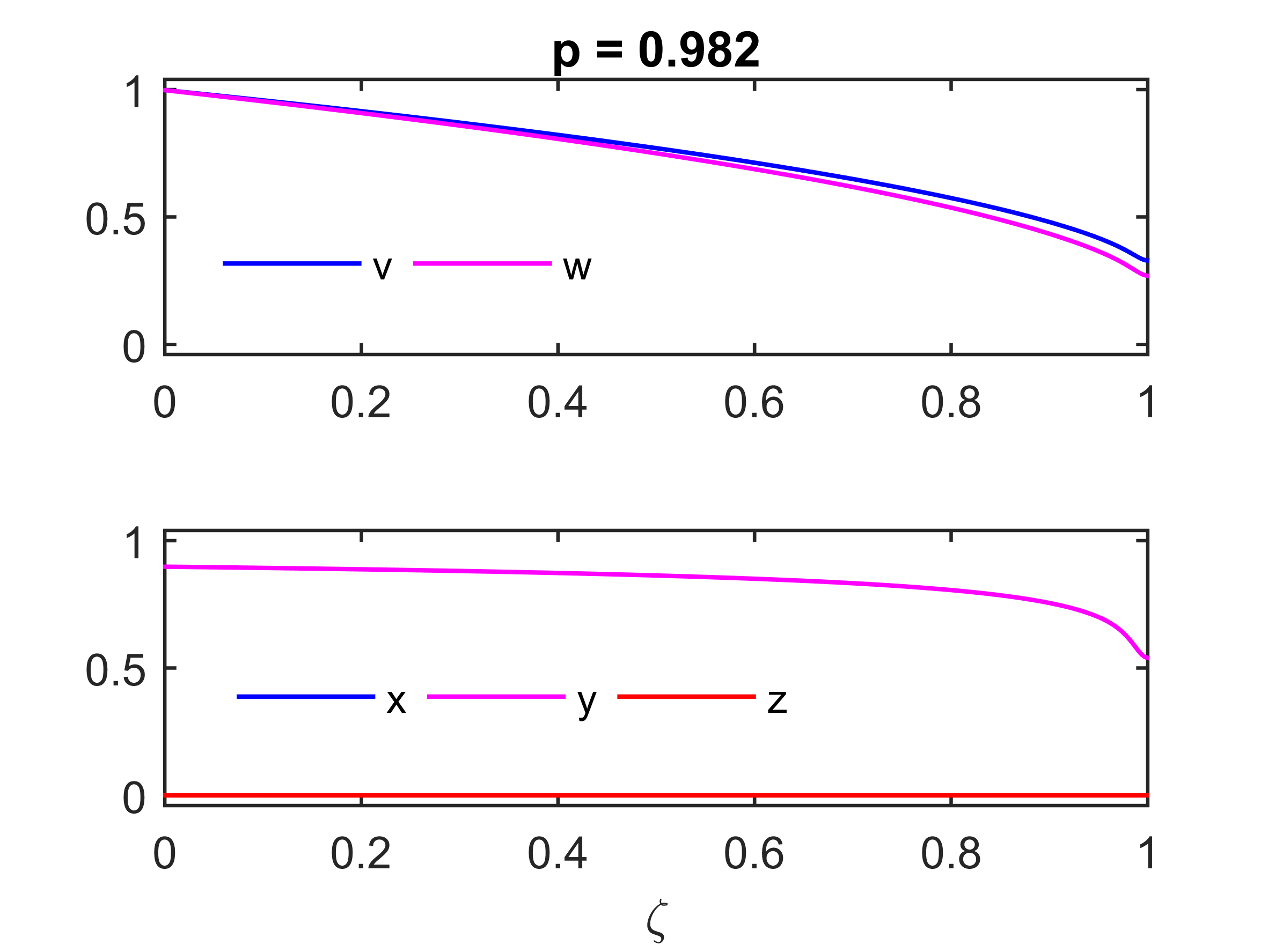}
	\caption{Stationary state of the system (\ref{fr}) at $p=0.982$. Top: relative concentration of reagents in the gas flow. Bottom: concentrations of surface species.}
	\label{fg_ss_982}
\end{figure}

At $p$ = 0.982, there are no more oscillations in the model and the solution of system (\ref{frfd}) tends to the stationary state shown in Figure \ref{fg_ss_982}. Now the surface of the catalyst is completely reduced and covered with x. The reaction proceeds only at the right (rear) edge of the plate.

The stationary solution at $p$ = 0.982 was continued to the left by parameter $p$ up to $p$ = 0.9815, and the solution remained stable. In the interval $0.8723<p<0.9815$, where the stationary solution undergoes radical reconstruction, using the methods of this work and our computational capabilities, it seems impossible to obtain a stationary solution of system (\ref{frfd}).

\subsubsection{The effect of the residence time $t_L$}

\begin{figure}[h]
	\centering
	\includegraphics{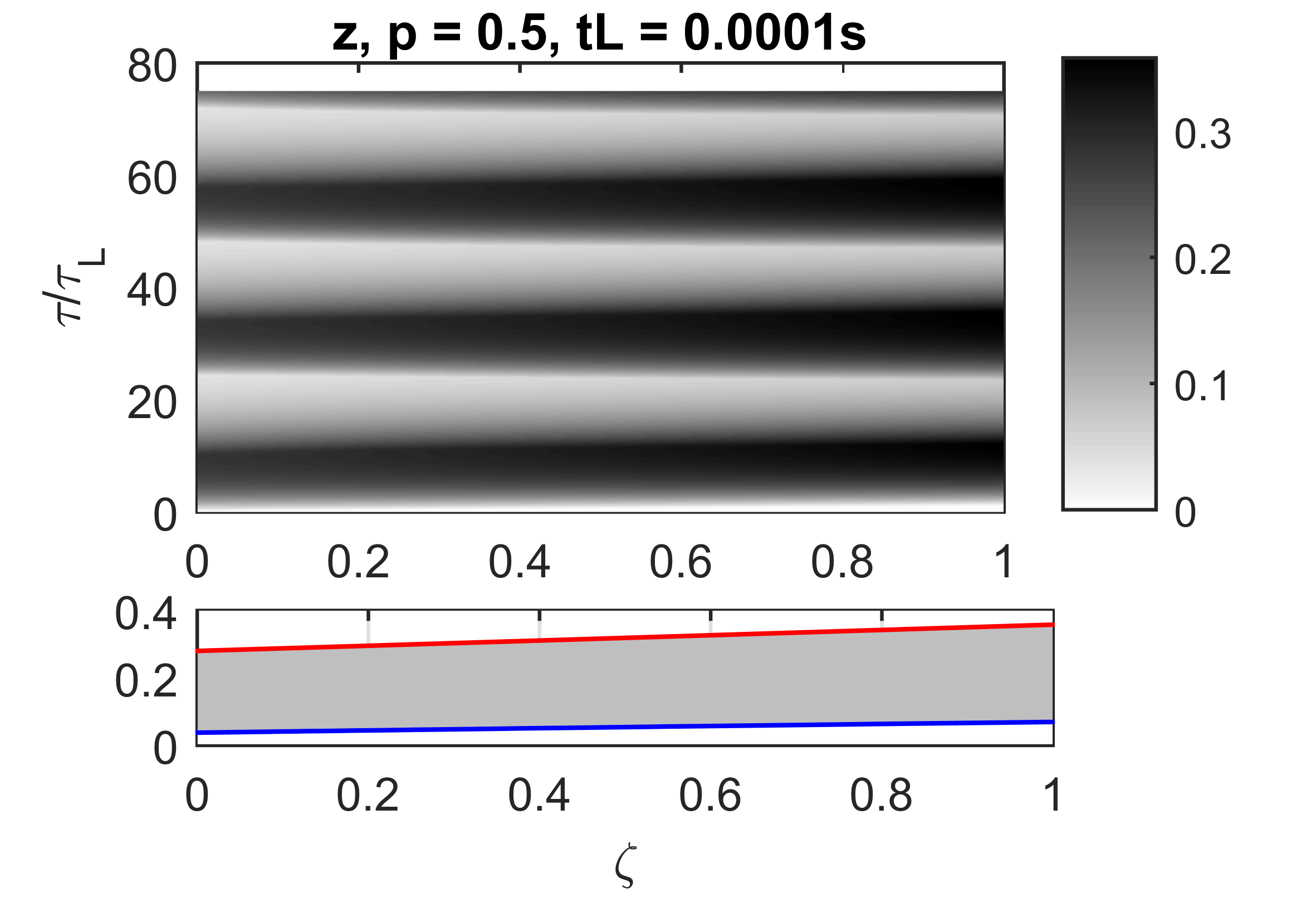}
	\caption{Oscillations of the oxidation degree $z$ at $p$ = 0.5 and $t_L$ = 0.0001s.}
	\label{fg_p05_t0001}
\end{figure}

The time $t_L = L/u$ during which the gas molecules in the flow pass over the surface of the catalyst determines the rate of replenishment of the mass of reagents to participate in the reaction. If this time is less than the characteristic reaction time, then the reaction proceeds at almost constant concentrations of reactants in the gas phase over the whole catalyst surface and the reaction can be described by a kinetic model. An example of oscillations at small $t_L$ is shown in Figure \ref{fg_p05_t0001}.

Figure \ref{fg_p05_t0001} represents the coherent oscillations of the oxidation degree (and other model variables) over the entire surface of the catalyst. Moreover, the amplitude and frequency of oscillations are close to the amplitude and frequency of oscillations obtained in the kinetic model.

\begin{figure}[h]
	\centering
	\includegraphics{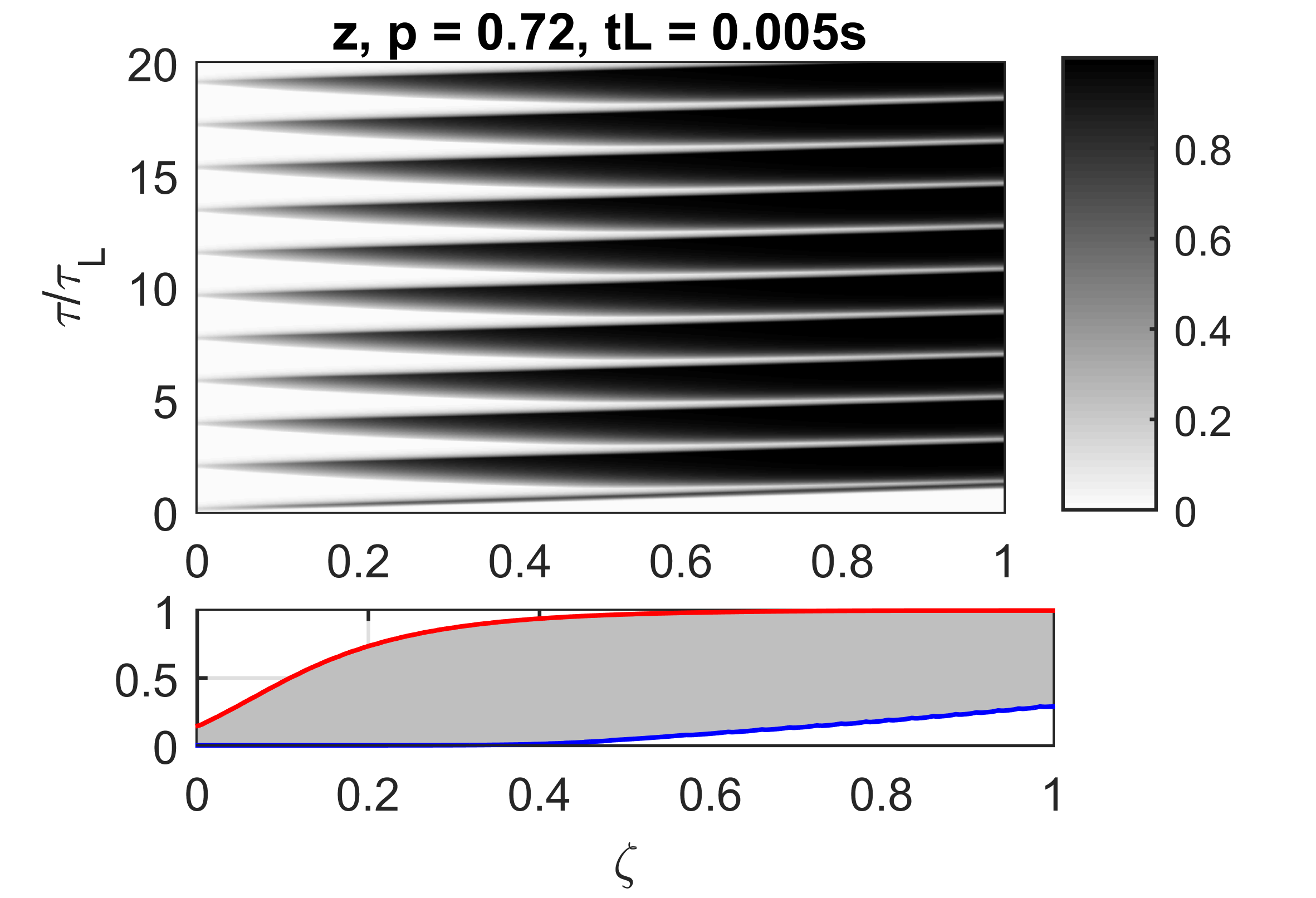}
	\caption{Moving fronts of oxidation and reduction at $p$ = 0.72, $t_L$ = 0.005 s.}
	\label{fg_p072_t005}
\end{figure}

With an increase in the time of flight $t_L$, the coherence of the oscillations is violated. Different parts of the surface along the gas flow are in significantly different conditions -- at different concentration of reagents in the gas phase. Alternating ``fronts'' of oxidation and reduction appear on the surface, moving in opposite directions. The oxidation front moves from the rear edge of the plate to the front, and the reduction front from the front to the back. An example of such fronts is shown in Figure \ref{fg_p072_t005}.

With a long time of flight, the flow does not have time to replenish the consumption of reagents for the reaction, and a non-uniform distribution of reagent concentrations along the catalyst surface occurs in the reactor. Increasing the time of flight leads to at least two effects. First, the range of values of the parameter $p$, in which oscillations exist, increases. Secondly, the width of the band on the catalyst surface decreases, on which oscillations in the concentrations of reagents are possible.

\begin{figure}[h]
	\centering
	\includegraphics{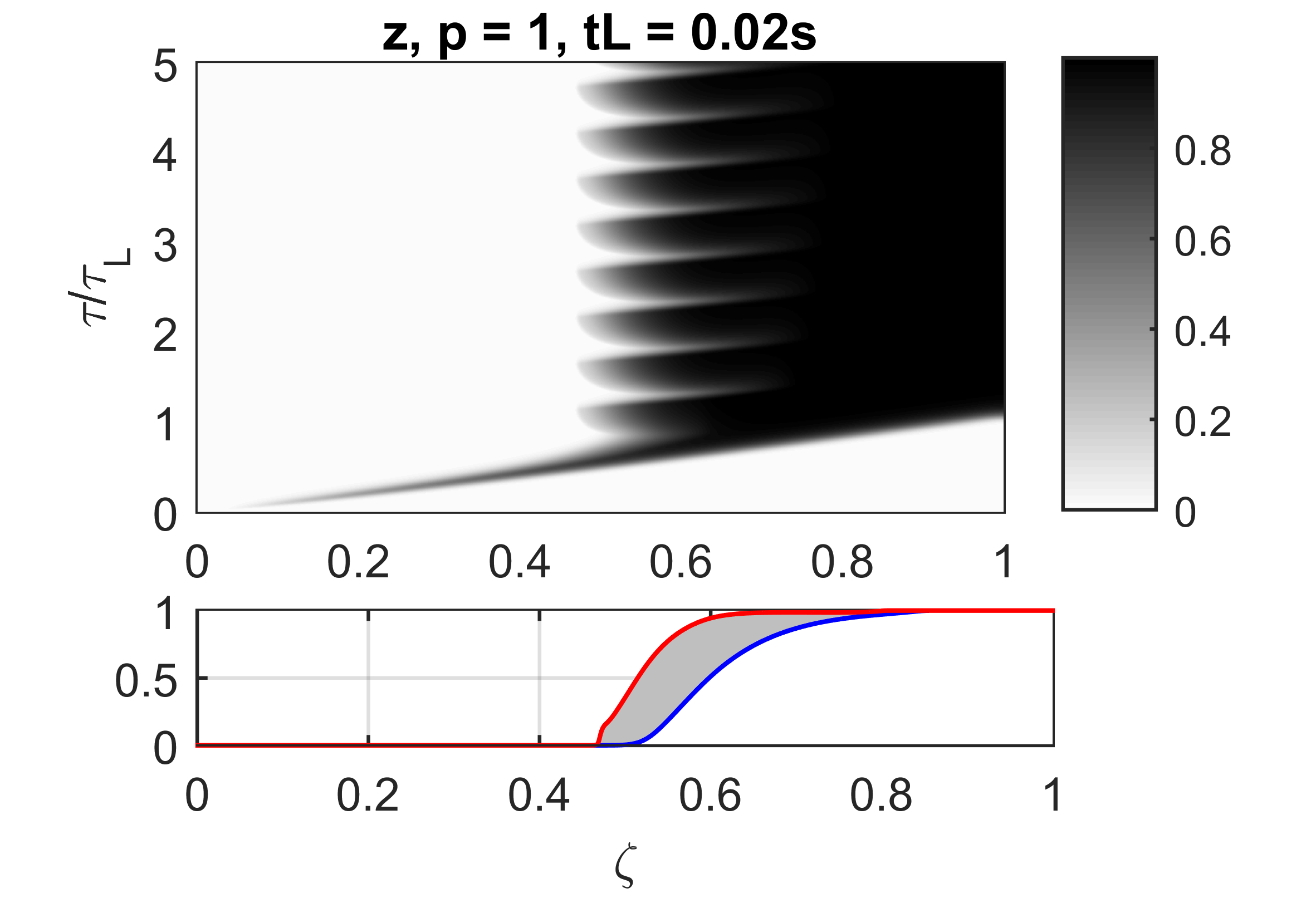}
	\caption{Localization of surface oscillations at $p$ = 1, $t_L$ = 0.02 s.}
	\label{fg_p1_t02}
\end{figure}

Oscillation localization is observed. Note that, by essence, Figures \ref{fg_os_446}-\ref{fg_os_981} show examples of such localization. As time $t_L$ increases, the localization becomes clearer. Figure \ref{fg_p1_t02} shows an example of localization at $t_L$ = 0.02 s.

\begin{figure}[h]
	\centering
	\includegraphics{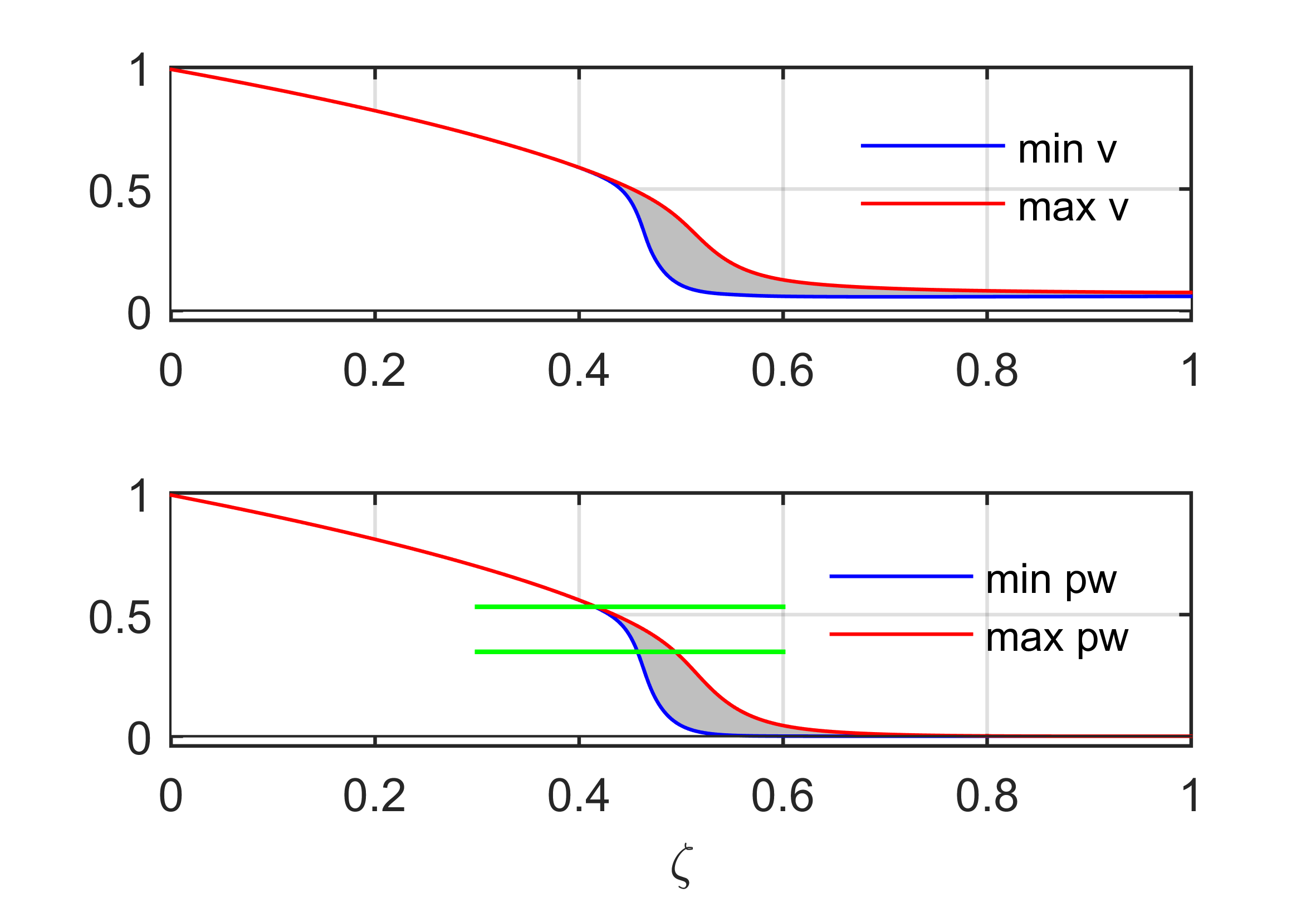}
	\caption{Localization of surface oscillations at $p$ = 1, $t_L$ = 0.02 s.}
	\label{fg_p1_t02_vw}
\end{figure}

Some details of localization are shown in Figure \ref{fg_p1_t02_vw}. This Figure shows graphs of the minimum and maximum values of the relative concentrations of reagents in the gas phase, depending on the coordinate $\zeta$. On the lower panel, green lines mark the boundaries $p_1$ and $p_2$ of the range of $p$ values, in which there are oscillations in the kinetic model.

\clearpage

It can be seen that oscillations on the catalyst surface arise when the graph of $pw$, which stands in place of $p$ in the kinetic model, crosses the upper boundary of this interval. The continuation of oscillations beyond the lower boundary of the interval $(p_1, p_2)$ is due to the gas flow. Thus, one can conclude that oscillations arise where condition $p_1<pw(\zeta,\tau)<p_2$  is satisfied on a fairly wide band of the catalyst surface.

\subsubsection{Impact of surface migration}

Recall that all previous results were obtained under the assumption that the surface migration coefficient $H$ is equal to the diffusion coefficient in the gas phase $D$. Here we give two examples of solving a system (\ref{frfd}) at $p=0.82$, $t_L = 0.01s$ and with a migration coefficient an order of magnitude greater and an order of magnitude less  than $D$. 

\begin{figure}[h]
	\centering
	\includegraphics{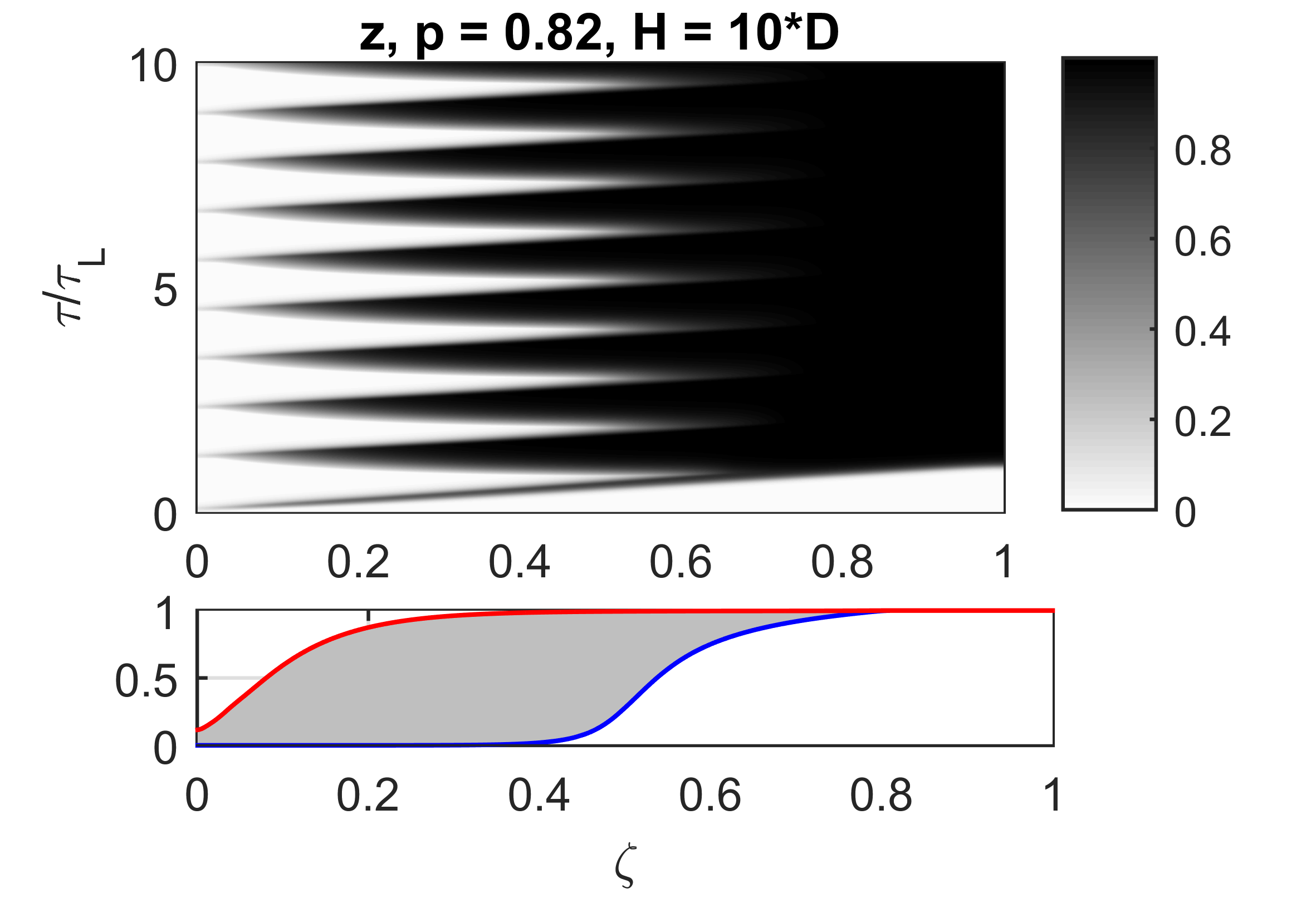}
	\caption{Space-time plot of the surface oxidation degree at $H=10D$.}
	\label{fg_p082_h10}
\end{figure}

\begin{figure}[h]
	\centering
	\includegraphics{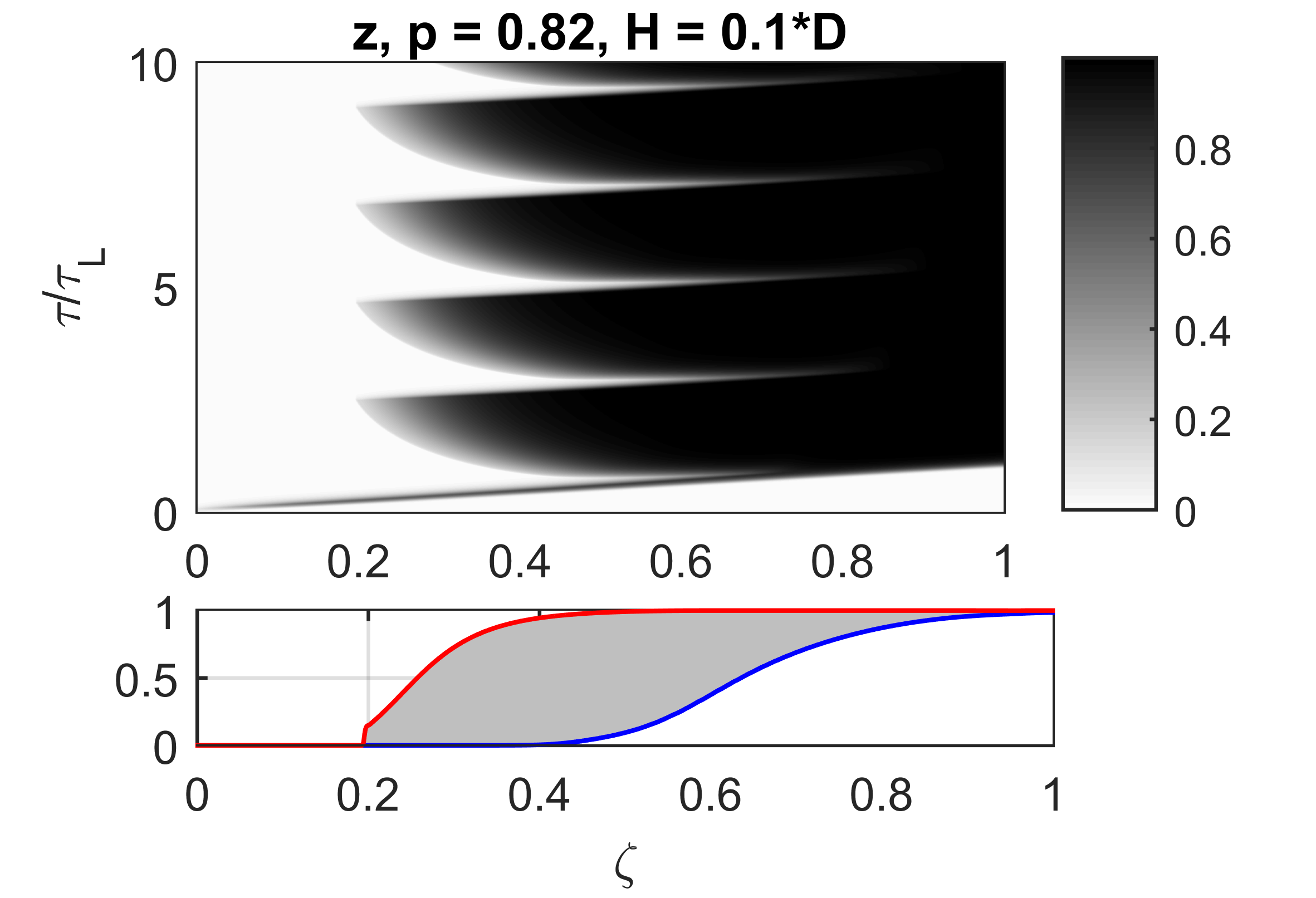}
	\caption{Space-time plot of the surface oxidation degree at $H=0.1D$.}
	\label{fg_p082_h01}
\end{figure}

Figure \ref{fg_p082_h10} shows the space-time plot of $z(\zeta,\tau)$ at $H = 10D$. The comparison with Figure \ref{fg_os_82}, which presents a similar picture for $H = D$, shows that with increasing $H$, the oscillation frequency increases and the oscillation band on the plate surface expands. This conclusion is supported by Figure \ref{fg_p082_h01}, which shows a similar plot for $H = 0.1D$. It can be seen that as $H$ decreases, the oscillation frequency and the oscillatory domain also decrease.

\begin{figure}[h]
	\centering
	\includegraphics{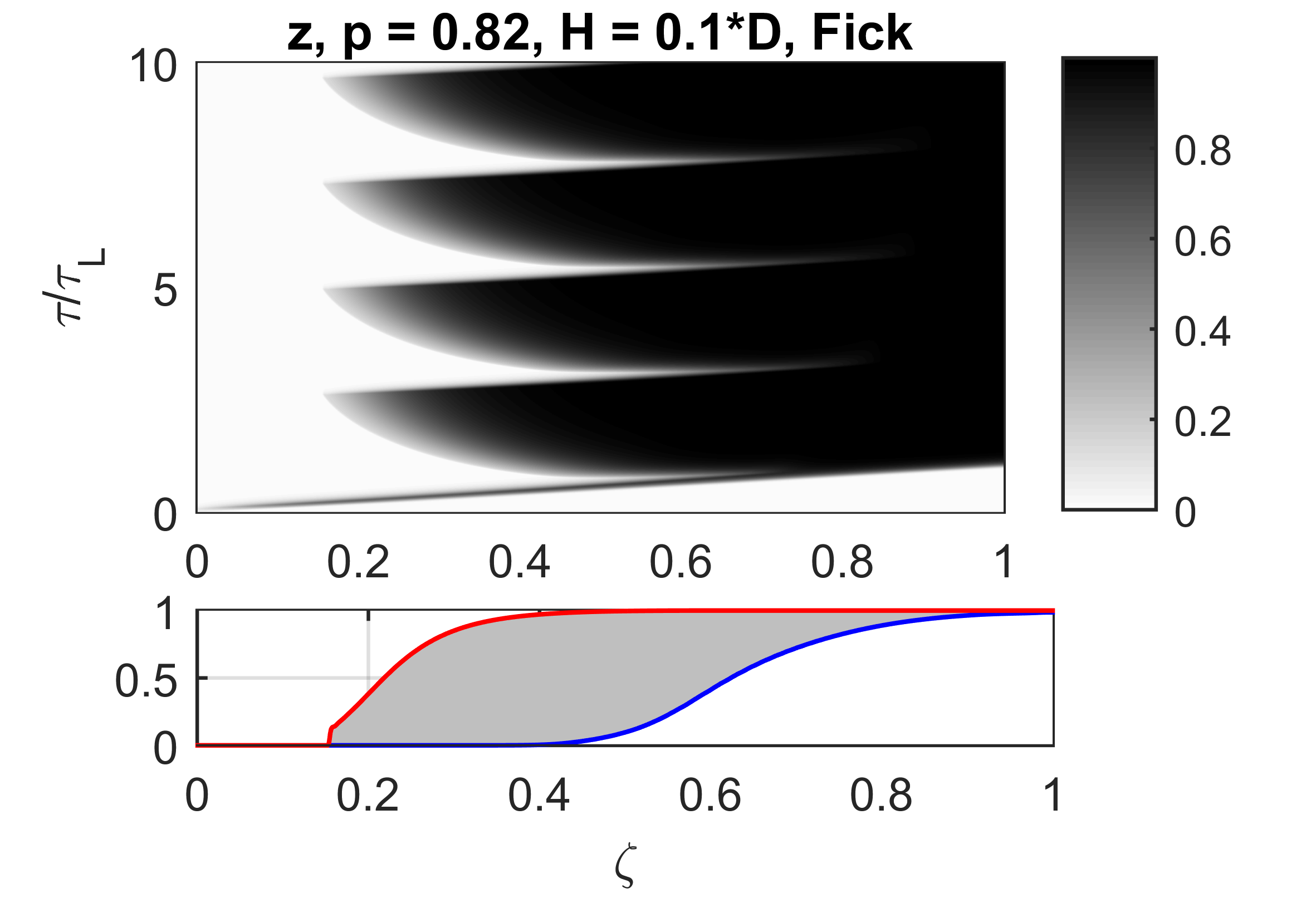}
	\caption{Space-time plot of the surface oxidation degree at $H=0.1D$ and classical Fick diffusion.}
	\label{fg_p082_h01_F}
\end{figure}

\clearpage

In this paper, the surface migration is described by a nonlinear equation (\ref{mig}). If one removes the nonlinear terms from this equation, then the classical Fick diffusion remains. For comparison, Figure \ref{fg_p082_h01_F} shows the space-time diagram for $z$ calculated with the classic diffusion. The pictures are very similar, but compared to the nonlinear migration, the frequency of oscillations here somewhat decreases and, on the contrary, the oscillation band on the catalyst surface expands. From a computational point of view, the scheme with the nonlinear migration is somewhat softer than the scheme with the classic diffusion -- in the first case, the computer time for solving the problem is somewhat less than in the second case.

\subsection{Moving fronts and reaction rate distribution}

\begin{figure}[h]
	\centering
	\includegraphics{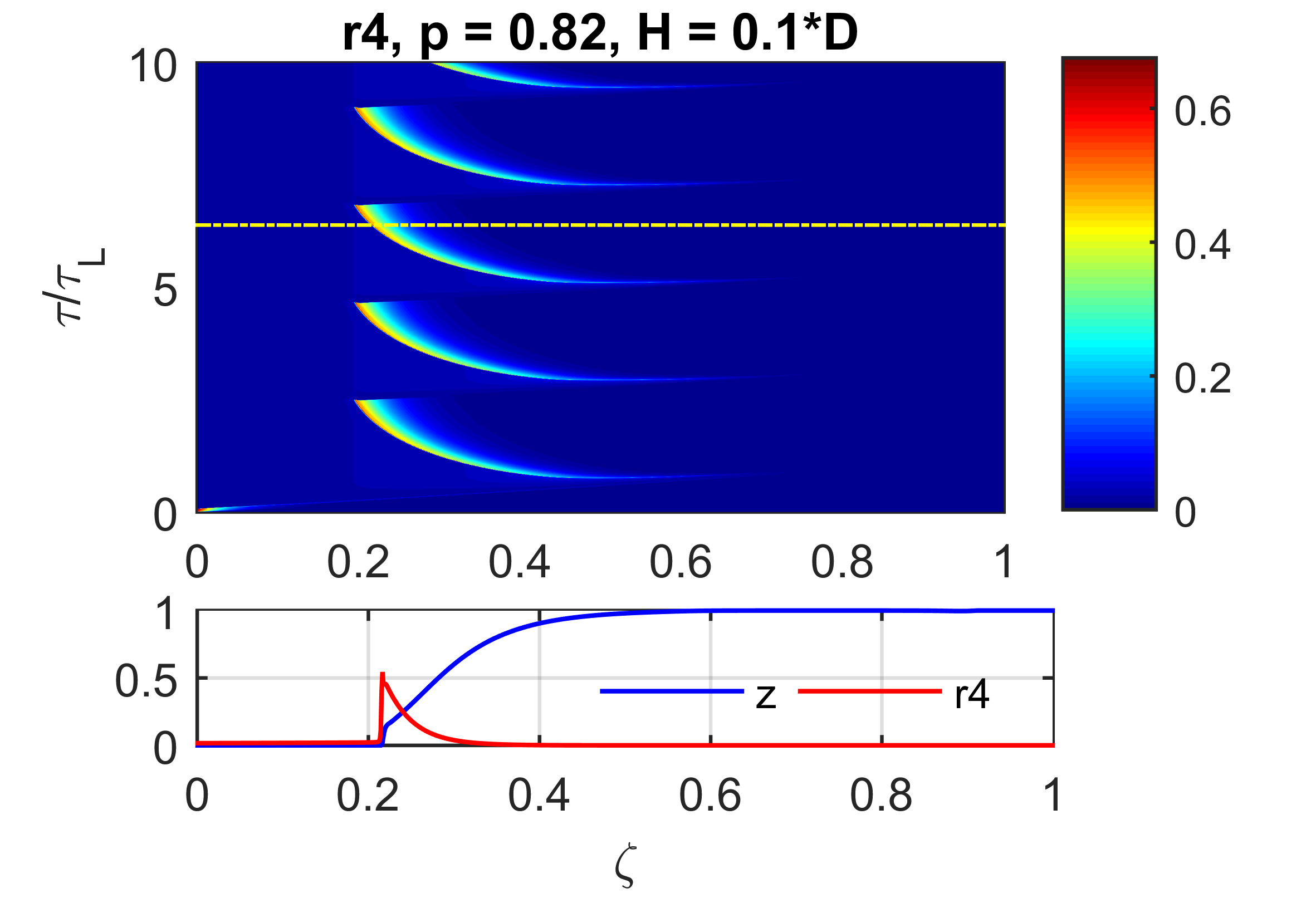}
	\caption{Top: space-time plot of the reaction rate $R_4=\kappa_4xy$. Bottom: the oxidation degree $z$ and reaction rate $R_4$ at the point in time marked with the horizontal yellow line in the top figure. $p = 0.82, H = 0.1D$.}
	\label{fg_p082_h01_rr}
\end{figure}

On the space-time sweeps of the oscillations shown above, the moving fronts of the oxidation and reduction of the catalyst are clearly visible. The oxidation front moves against the gas flow, reaches the edge of the plate or some point inside the plate, and stops. Then, from the place where the oxidation front stops, the reduction front begins to move, which moves in the opposite direction -- along the flow. Note that the surface reaction x + y $\to$ XY occurs mainly at the oxidation front, as shown in Figure \ref{fg_p082_h01_rr}.

Figure \ref{fg_p082_h01_rr} shows the space-time plot of the reaction rate $R_4$ for the solution of the system shown in Figure \ref{fg_p082_h01}, which of all the shown solutions represents the reaction rate distribution most expressively. The oxidation front advances on the reduced surface. The reduced surface is covered with adsorbed molecules of the reducing agent y, there is very little oxidizing agent x on it, and the x + y reaction practically does not occur. At the oxidation front, the concentration y decreases due to the reduction reaction y + Mx. The possibility opens up for the adsorption of the oxidizing agent X and, consequently, the reaction x + y. With advancement into the oxidized surface behind the oxidation front, the reaction decays due to a decrease in the concentrations of y on the surface and Y in the gas phase.

\subsection{Low frequency oscillations}

In this section, we shall take $\kappa_5=\kappa_6=10^{-6}$, so the oscillation frequency decreases by 4 orders of magnitude compared to the previous section and becomes about $10^{-2}$ s$^{-1}$. At the same time, the bifurcation diagram of the kinetic model with such constant values differs little from Figure \ref{fg_km_ss}. In particular, now the limit cycle is observed on the interval 0.339 $<p<$ 0.532, while in the previous case it was 0.347 $<p<$ 0.532.

\begin{figure}[h]
	\centering
	\includegraphics{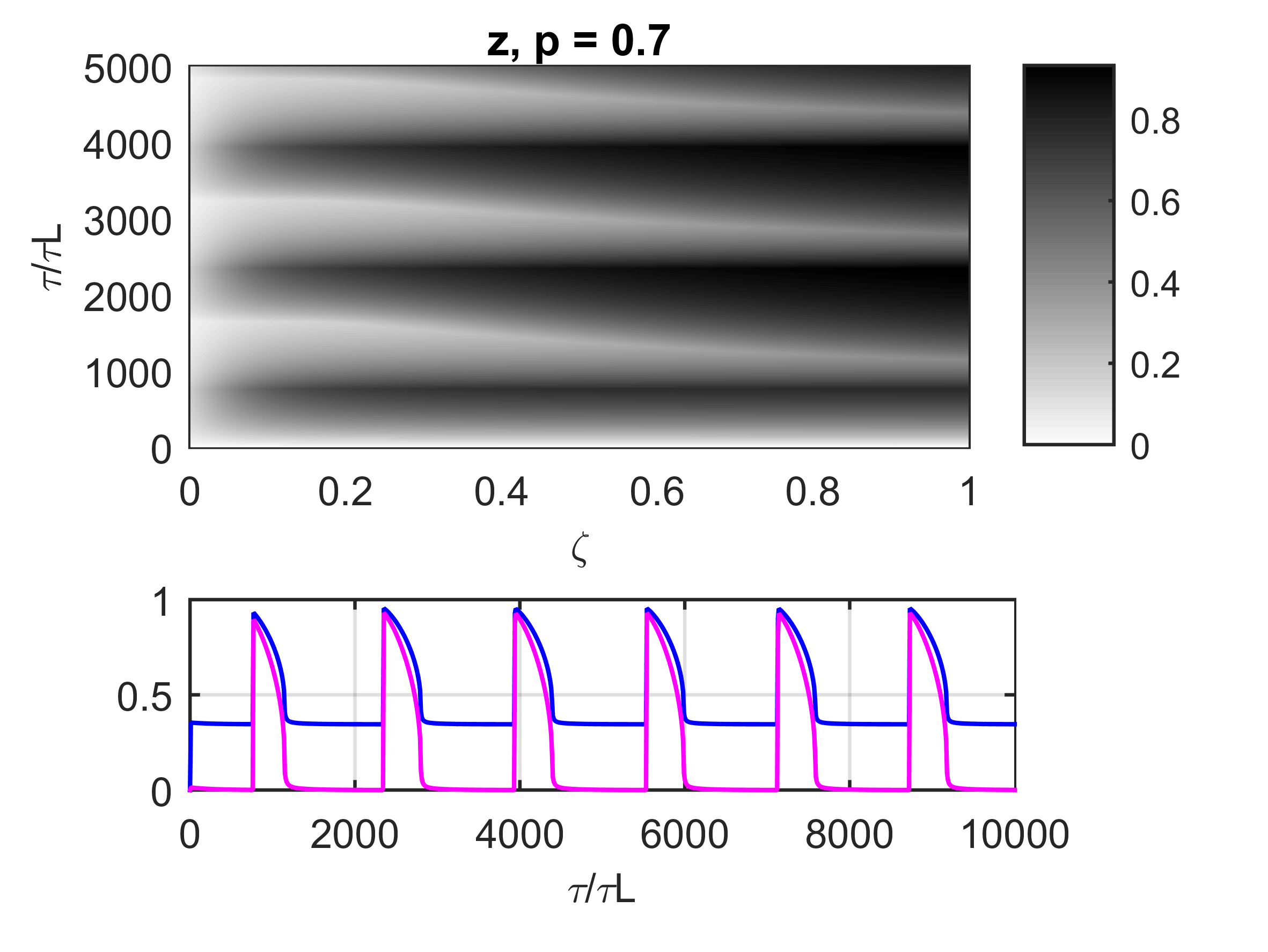}
	\caption{Top: space-time plot of $z$ oscillations. Bottom: time series of $v(1,\tau)$ -- blue line, $w(1,\tau)$ -- magenta line, at $\kappa_5=\kappa_6 = 10^{-6}$, $p$ = 0.7.}
	\label{fg_p07_lf}
\end{figure}

\begin{figure}[h]
	\centering
	\includegraphics{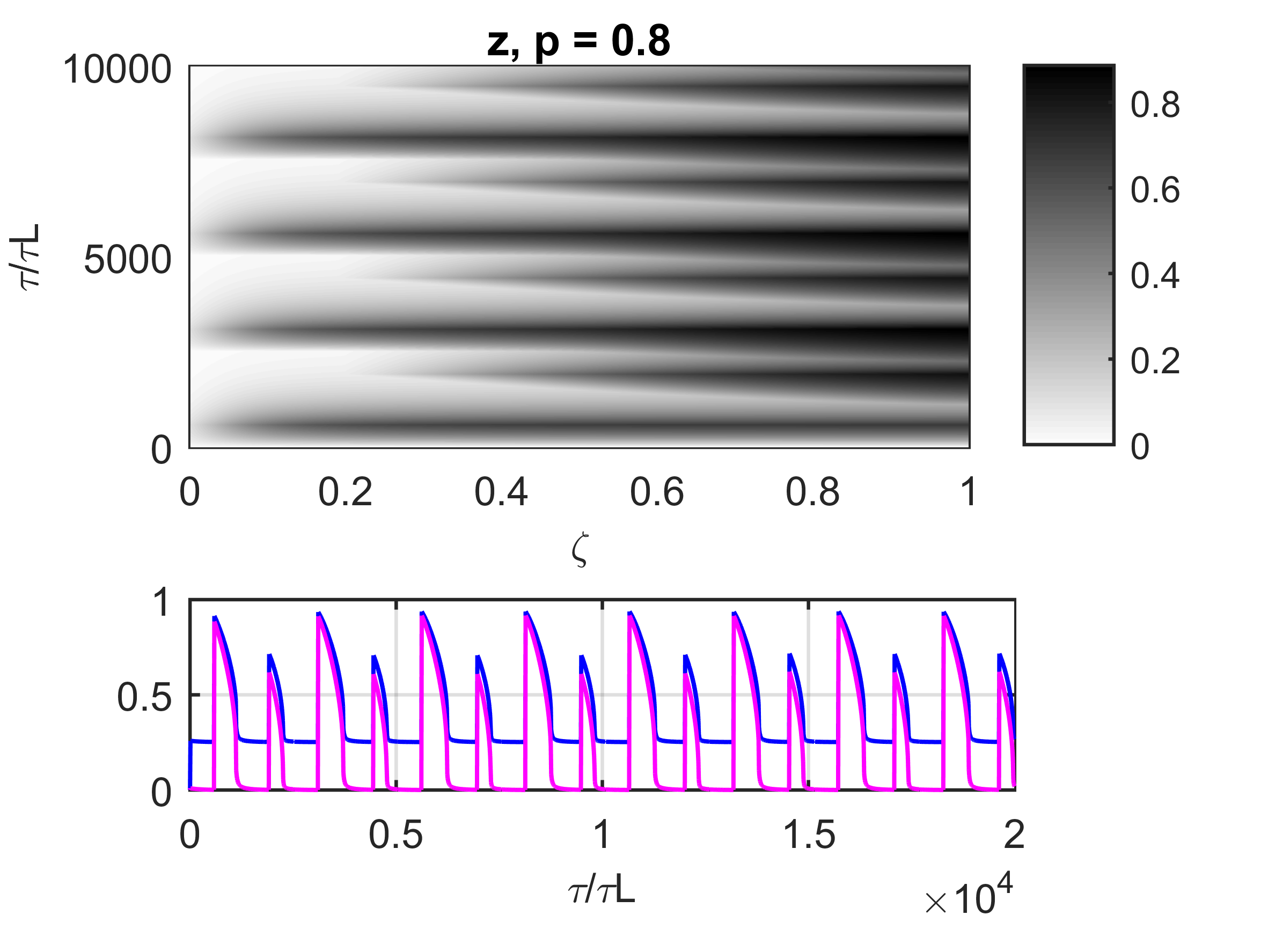}
	\caption{The same as in Fig. \ref{fg_p07_lf} for $p$ = 0.8.}
	\label{fg_p08_lf}
\end{figure}

\begin{figure}[h]
	\centering
	\includegraphics{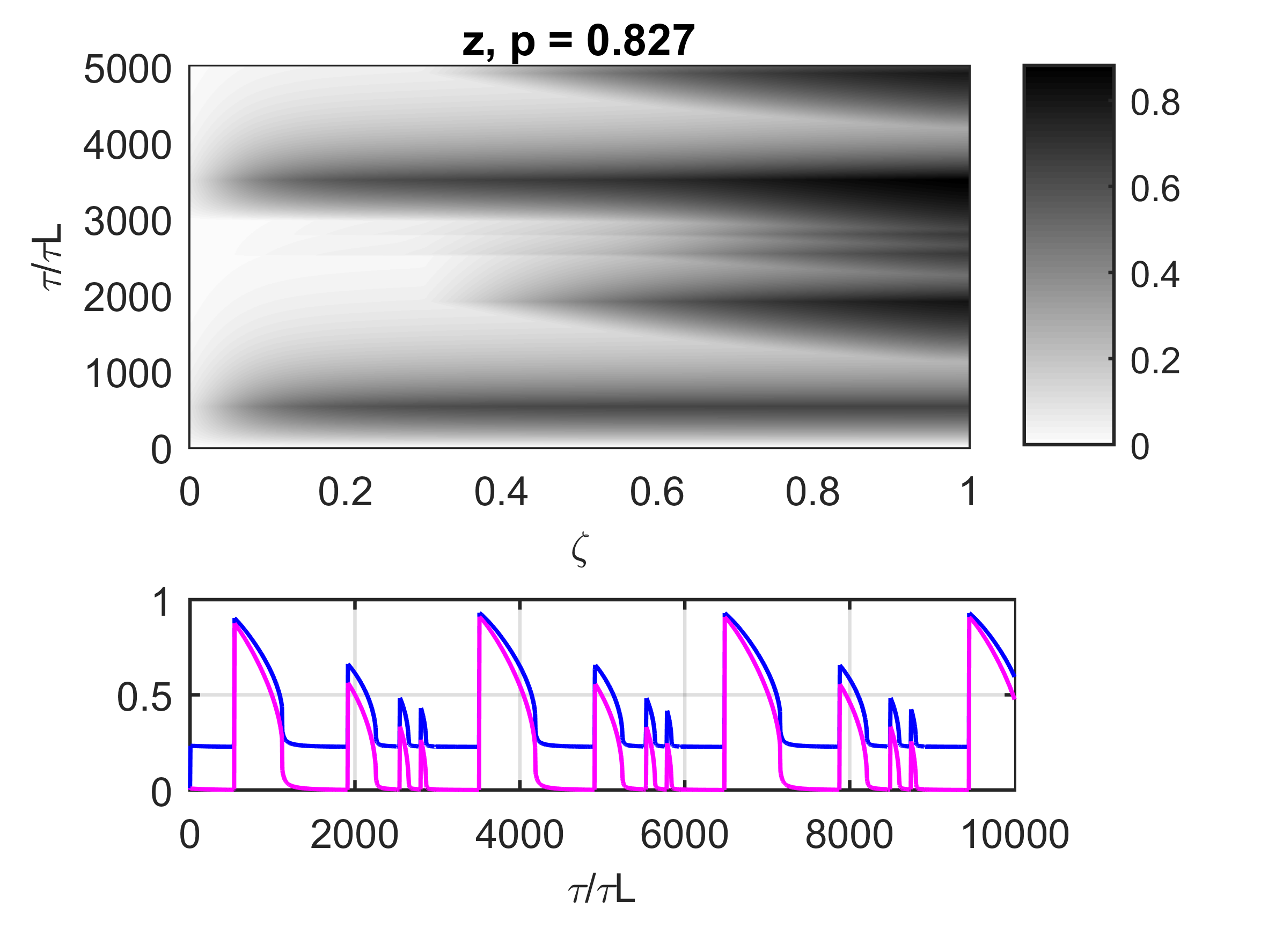}
	\caption{The same as in Fig. \ref{fg_p07_lf} for p = 0.827.}
	\label{fg_p0827_lf}
\end{figure}

%\clearpage

\begin{figure}[h]
	\centering
	\includegraphics{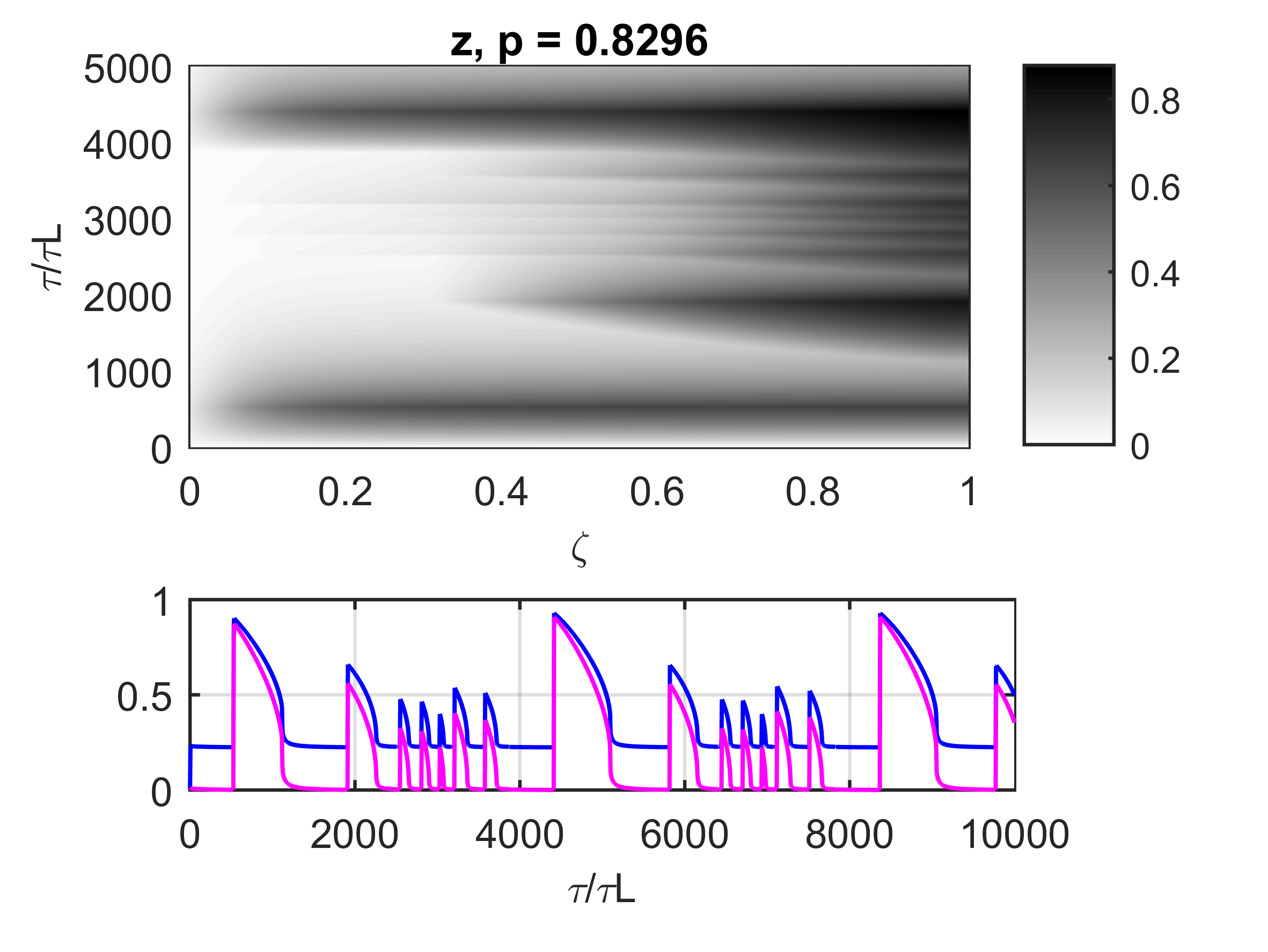}
	\caption{The same as in Fig. \ref{fg_p07_lf} for p = 0.829.}
	\label{fg_p08296_lf}
\end{figure}

\begin{figure}[h]
	\centering
	\includegraphics{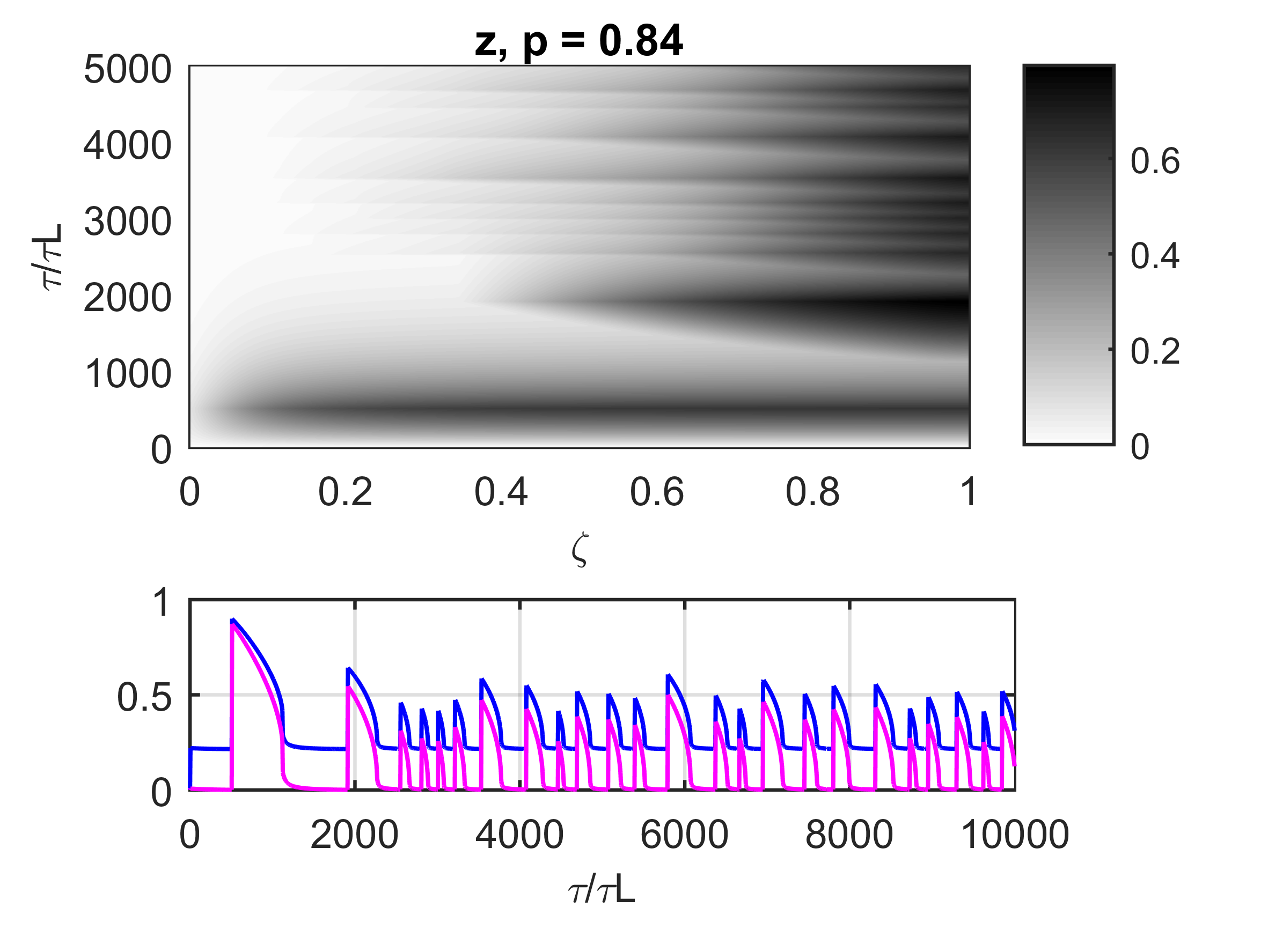}
	\caption{The same as in Fig. \ref{fg_p07_lf} for p = 0.84.}
	\label{fg_p084_lf}
\end{figure}

\clearpage

The high-frequency oscillations for the considered values of the parameters have the form of (simple, one-turn) regular oscillations. The form of low-frequency oscillations changes with a change in the value of the parameters, in particular, the parameter $p$. Oscillations can become more complex, and even changes are observed that resemble Feigenbaum scenario of transition to deterministic chaos. An example of such complication is shown in Figures \ref{fg_p07_lf}-\ref{fg_p084_lf}. These figures, like the previous ones, consist of two parts. The upper part shows the space-time plot of the degree of oxidation of the catalyst surface $z(\zeta,\tau)$, and the lower part shows the time dependence of the relative concentration of reactants in the gas phase at the outlet of the reactor.

As in the high frequency case, low frequency oscillations arise at a certain value of $p$ at the frontal (left) edge of the plate and, with increasing $p$, quickly propagate to the entire surface of the plate. The space-time plot in Figure \ref{fg_p07_lf} ($p$ = 0.7) shows alternating moving fronts of oxidation and reduction. Moreover, both fronts propagate from right to left, from the rear edge of the plate to the front.

As the parameter $p$ approaches the right edge of the interval of existence of the oscillatory solution in the flow reactor model, the oscillations quickly become more complex from the "simple" oscillations shown in Figure \ref{fg_p07_lf} to the irregular oscillations shown in Figure \ref{fg_p084_lf}. Due to limited computational capabilities, we cannot present a complete picture of how the solution changes with increasing parameter $p$, so we present several examples of solutions. First, a “period-doubling” occurs -- each second oxidation front does not reach the left edge of the plate. An example of a "doubled period" oscillation is shown in Figure \ref{fg_p08_lf} ($p$ = 0.8). Then the period doubles again, -- Figure \ref{fg_p0827_lf} ($p$ = 0.827). The next example, Figure \ref{fg_p08296_lf} ($p$ = 0.829), is not part of the period-doubling sequence, but shows a "7-cycle". Note, however, that the resemblance to the Feigenbaum scenario here is purely superficial. The process of complication of oscillations and the final irregular oscillations require a detailed analysis, which is beyond the scope of this work.

\begin{figure}[h]
	\centering
	\includegraphics{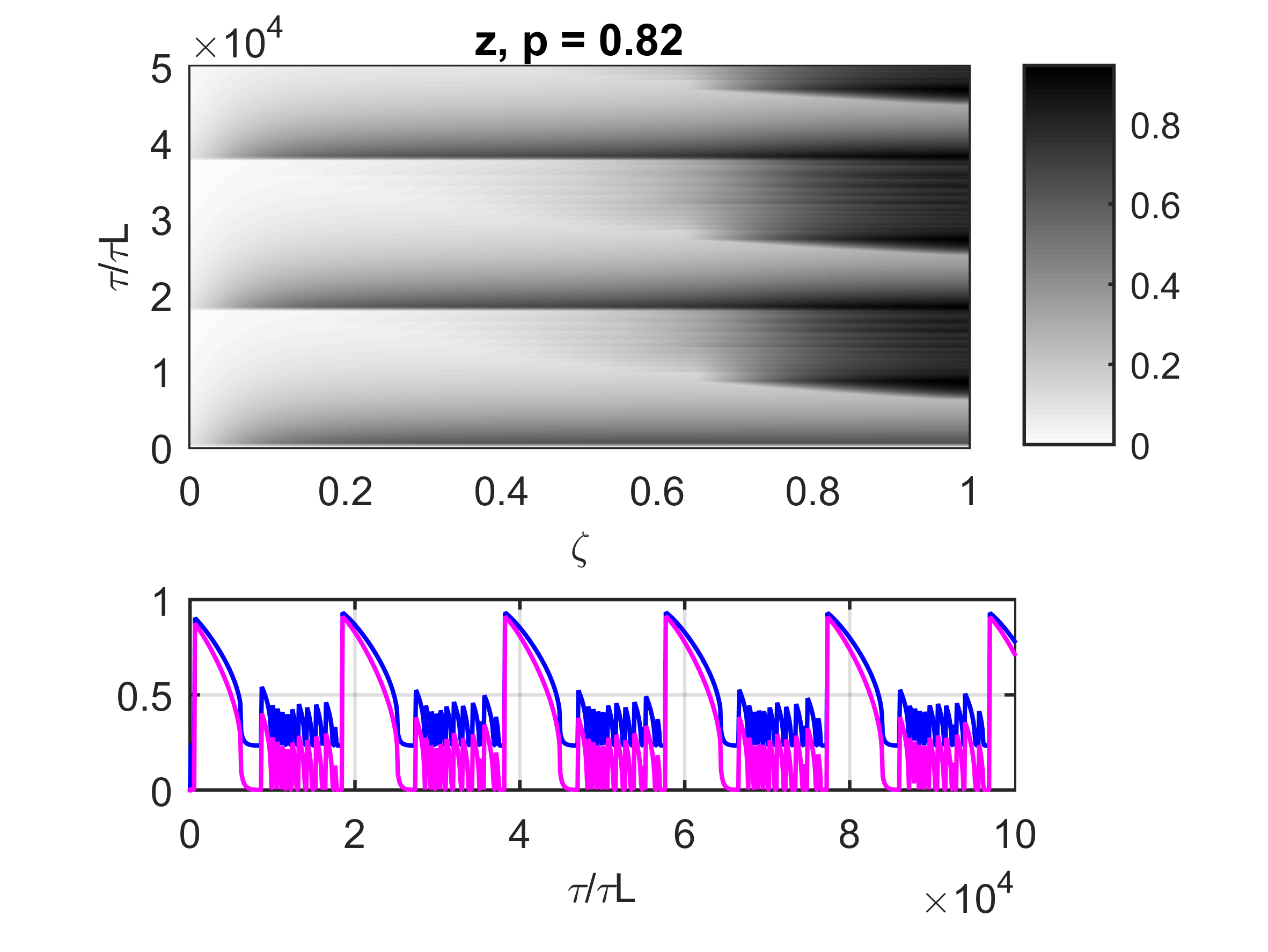}
	\caption{The same as in Fig. \ref{fg_p07_lf} for p = 0.82, $r_5=10^{-6}$, $r_6=10^{-7}$ .}
	\label{fg_p082_r01_lf}
\end{figure}

In this work, we limited ourselves to the assumption that the rate of oxidation of the  catalyst $\kappa_5$ is equal to the rate of its reduction $\kappa_6$. However, the relation $\kappa_5/\kappa_6$ plays a significant role in the dynamics of the model, which we do not discuss here. The only example shown in Figure \ref{fg_p082_r01_lf} demonstrates the appearance of complex mixed mode oscillations for different values of oxidation and reduction rates, namely for $\kappa_5=10^{-6}$, $\kappa_6=10^{-7}$, at p = 0.82.

\section{Concluding remarks}

The results obtained in this work show that complex regimes can be realized in a flow reactor for the reaction mechanism described by the simplest kinetic model. It was clearly demonstrated that the region of oscillations in the parameter space of the distributed model of the flow reactor is much larger than in the point kinetic model. The spatial-temporal behavior of the distributed system (\ref{fr}) is highly dependent upon the frequency of oscillations in the point kinetic model.

High-frequency oscillations with a period of not more than several times of flight $t_L$ of the gas flow over the catalyst surface have a regular shape over the entire range of values of the parameter $p$, on which oscillations exist. For small values of $p$, the solutions of the system tend over time to a stationary state in which the catalyst surface is oxidized. The reaction proceeds only at the leading edge of the catalyst plate. As $p$ increases, oscillations arise when the inequality $pw(\zeta,\tau) > p_1$ is satisfied in a sufficiently wide band of $\zeta$ near the left edge of the plate ($\zeta=0$). In this case, the Hopf bifurcation occurs in the discrete system (\ref{frfd}). The bifurcation occurs at a larger value of $p$, then in the kinetic model, because due to the Dankwerts condition, $w(0,\tau) < 1$. At further increase in $p$, the oscillation band on the plate expands in the direction of the flow and shifts towards the back edge of the plate. The width of the oscillation band depends on the time of flight $t_L$ (or on the flow velocity $u$), -- the larger $t_L$, the narrower the band.

The oscillations arising in the band of fulfillment of inequality $p_1 < pw(\zeta,\tau) < p_2$ are transferred by the gas flow to a wider band and even to the entire plate. Therefore, the observed region of oscillations on the plate is much wider than the band on which they originate.

Low-frequency oscillations with a period of several thousand times of flight $t_L$  demonstrate, in contrast to high-frequency ones, diverse dynamics. In particular, as the parameter $p$ approaches the right end of the oscillation interval, oscillations become more complicated. In this work, the complication of oscillations is demonstrated by the example of the concentration of reagents at the reactor outlet, which can be measured experimentally. As $p$ increases, the oscillations take the form of mixed mode oscillations with a gradual increase in the number of maxima in the period. When $p$ is close to the boundary of the oscillatory interval, the oscillations become aperiodic.

%{00}

\end{document}